\documentclass[]{interact}

\usepackage{epstopdf}
\usepackage{subcaption}
\usepackage{float}
\usepackage{comment}
\usepackage{optidef}
\usepackage{tikz}
\usepackage{dsfont}
\usepackage{graphicx}

\usepackage[natbibapa,nodoi]{apacite}
\theoremstyle{plain}

\theoremstyle{definition}

\theoremstyle{remark}

\begin{document}


\title{A Modified CTGAN-Plus-Features Based Method for Optimal Asset Allocation}

\author{
\name{José-Manuel Peña\textsuperscript{a}\thanks{Contact email: research@fintual.com}, Fernando Suárez\textsuperscript{a}, Omar Larré\textsuperscript{a}, Domingo Ramírez\textsuperscript{a}, Arturo Cifuentes\textsuperscript{b} }
\affil{\textsuperscript{a} Fintual Administradora General de Fondos S.A. Santiago, Chile. Fintual, Inc. \\
\textsuperscript{b} Clapes UC, Pontificia Universidad Católica de Chile, Santiago, Chile.  } 
}

\maketitle

\begin{abstract}

We propose a new approach to portfolio optimization that utilizes a unique combination of synthetic data generation and a CVaR-constraint. We formulate  the portfolio optimization problem as  an asset allocation problem  in which each asset class is accessed through a passive (index) fund.  The asset-class weights are determined by solving an optimization problem which includes a CVaR-constraint.  The optimization is carried out by means of a Modified CTGAN algorithm which incorporates features (contextual information) and is used to generate synthetic return  scenarios, which, in turn, are    fed into the  optimization engine.  For contextual information we rely on several points along the U.S. Treasury yield curve.  The merits of this approach are demonstrated with an example based on ten asset classes (covering stocks, bonds, and commodities) over a fourteen-and-half year period (January 2008-June 2022).  We also show that the synthetic generation process is able to capture well the key characteristics of the original data, and the optimization scheme results in portfolios that exhibit satisfactory out-of-sample performance.  We also show that this approach outperforms  the conventional equal-weights (1/N) asset allocation strategy and other optimization formulations based on historical data only.

\end{abstract}


\section{Motivation and Previous Work}

The portfolio selection problem—how to spread a given budget among several investment options—is probably one of the oldest problems in applied finance.  Until 1952, when Harry Markowitz published his famous portfolio selection paper, the issue was tackled with a mix of gut feeling, intuition, and whatever could pass for common sense at the moment.  A distinctive feature of these approaches was that they were, in general, qualitative in nature.

Markowitz’s pioneering work showed that the portfolio selection problem was in essence an optimization problem that could be stated within the context of a well-defined mathematical framework \citep{markowitz1952}.  The key ideas behind this framework (e.g., the importance of diversification, the tradeoff between risk and return, and the efficient frontier) have survived well the test of time.  Not only that, Markowitz’s paper triggered a voluminous amount of research on this topic that was quantitative in nature, marking a significant departure from the past.

However, notwithstanding the merits of Markowitz’s approach (also known as mean-variance or MV portfolios), its implementation has been problematic.  First, estimating the coefficients of the correlation-of-returns matrix—the essential backbone of the MV formulation—is a problem that still lacks a practical solution. For example, \citet{demiguel2009optimal} concluded that in the case of a portfolio of 25 assets, estimating the entries of the correlation matrix with an acceptable level of accuracy would require more than 200 years of monthly data.  A second drawback of Markowitz’s formulation, more conceptual than operational, is that it relies on the standard deviation of returns to describe risk. However, the standard deviation, since its focuses on dispersion, is not a good proxy for risk for it really captures uncertainty—a subtle but significant difference (\citet{friedman2014risky}).  Anyhow, the fact of the matter is that during the second part of the previous century most research efforts were aimed at devising practical strategies to implement the MV formulation.  Needless to say, success in these efforts has been mixed at best and  these days most practitioners have moved beyond the original MV formulation, which only remains popular within some outdated academic circles.  \citet{KOLM2014356} summarize well the challenges associated with the implementation of Markowitz's approach. \citet{pagnocelli2022} provide a brief overview of the different techniques that have attempted to reconcile the implementation of the MV formulation with reality.

John Bogle, who founded the Vanguard Group (an asset management company) and is recognized as the father of index investing, is another pioneer whose main idea was revolutionary at the time and remains influential until  today.  In 1975 he introduced a concept known as passive investment.  He thought that a fund whose goal was to beat the market would necessarily have high costs, and hence investors would be better served by a low-cost fund that would simply mimic the market by replicating a relevant index \citep{bogle2018stay, thune_2022}.  This innovation, highly controversial at the time, has been validated by empirical evidence as study-after-study has demonstrated that trying to beat the market (in the context of liquid and public markets) is a fool’s errand (e.g., \citet{sharpe1991arithmetic, walden2015active, elton2019passive, fahling2019active}).  But Bogle’s idea had another important ramification that made the portfolio selection problem more tractable: it shifted the emphasis from asset selection to asset allocation.  More to the point, before the existence of index funds, an investor who wanted exposure to, say, the U.S. stock market —leaving aside the shortcomings of the MV formulation for a moment— faced an insurmountable large optimization problem (at least 500 choices if one restricts the feasible set to the stocks in the S\&P 500 index).  Today, the same investor can gain exposure to a much more diversified portfolio— for example, a portfolio made up of U.S. stocks, emerging markets stocks, high yield bonds, and commodities— simply by choosing an index fund in each of these markets and concentrating then in estimating the proper asset allocation percentages.  In short, a much smaller optimization problem (\citet{amenc2001s, ibbotson2010, GUTIERREZ2019134}).

In any event, this switch from asset selection to asset allocation, plus a number of innovations that emerged at the end of the last century and have gained wide acceptance in recent years, have changed the portfolio selection landscape in important ways.  Among these innovations we identify the following:

\begin{enumerate}
\item The Conditional-Value-at-Risk or CVaR has established itself as the risk metric of choice.  A key advantage is that it captures much better than the standard deviation the so-called tail risk (the danger of extreme events).  A second advantage is that by focusing on losses rather than volatility of returns is better aligned with the way investors express their risk preferences (\cite{rockafellar2000optimization}, \citet{rockafellar2002conditional}). A third advantage is that in the case of the discretization and linearization of the portfolio optimization problem, as we will see in the following section, the CVaR places no restrictions on the type of probability distribution that can be used to model the returns.

\item The benefits of relying on synthetic data to simulate realistic scenarios is crucial for solving stochastic optimization problems such as the one described by Markowitz.  As mentioned by \citet{doprado}, a financial modeler looking at past returns data, for example, only sees the outcome from a single realized path (one returns time series) but remains at a loss regarding the stochastic (data generating) process behind such time series.  Additionally, any effort aimed at generating realistic synthetic data must capture the actual marginal and joint distributions of the data, that is, all the other possible returns time histories that could have occurred but were not observed.  Fortunately, recent advances in neural networks and machine learning—for example, an algorithm known as Generative Adversarial Networks or GAN—has proven effective to this end in a number of applications \citep{goodfellow2014}.  Moreover, a number of authors have explored the use of GAN-based algorithms in portfolio optimization problems, albeit, within the scope of a framework different than the one discussed in this paper (e.g., \citet{lu2022autoencoding, pun2020financial, takahashi2019modeling, mariani2019}).  \citet{Lommers67}.    \citet{eckerli2021generative} provide a very good overview of the challenges and opportunities faced by machine learning in general and GANs in particular when applied to financial research.


\item There is a consensus among practitioners that the joint behavior of a group of assets can fluctuate between discrete states, known as market regimes, that represent different economic environments (\citet{hamilton1988, hamilton1989, schaller}). Considering this observation, realistic synthetic data generators (SDGs) must be able to account for this phenomenon.  In other words, they must be able to generate data belonging to different market regimes, according to  a multi-mode random process.

\item The incorporation of features (contextual information) to the formulation of many optimization problems has introduced important advantages.  For example, \citet{ban2019big} showed that by adding features to the classical newsvendor problem resulted in solutions with much better out-of-sample performance compared to more traditional approaches.  Other authors have also validated the effectiveness of incorporating features to other  optimization problems (e.g., \citet{hu2022fast, bertsimas2020predictive, chen2022statistical, see2010robust}).

\end{enumerate}

With that as background, our aim is to propose a method to tackle the portfolio selection problem based on an asset allocation approach.  Specifically, we assume that our investor has a medium- to long-term horizon and has access to a number of liquid and public markets in which he/she will participate via an index fund.  Thus, the problem reduces to estimating the appropriate portfolio weights assuming that the rebalancing is not done very frequently.  Frequently, of course, is a term subject to interpretation.  In this study we assume that the rebalancing is done once a year.  Rebalancing daily, weekly, or even monthly, would clearly defeat the purpose of passive investing while creating excessive trading costs, that ultimately could affect performance.

Our approach is based on a  Markowitz-inspired framework but with a CVaR-based risk constraint instead.  More importantly, we rely on synthetic returns data generated with a Modified Conditional GAN approach which we enhance with contextual information (in our case, the U.S. Treasury yield curve).  In a sense, our approach follows the spirit of \citet{pagnocelli2022}, but it differs in several important ways and brings with it important advantages—including performance—a topic we discuss in more detail later in this paper.  In summary, our goals are twofold.  First, we seek to propose  an effective synthetic data generation algorithm; and second, we seek to combine such algorithm with contextual information to propose an asset allocation method, that should yield, ideally, acceptable out-of-sample performance.

In the next  section we formulate the problem at hand more precisely, then we describe in detail the synthetic data generating process, and we follow  with a  numerical example.  The final section presents the conclusions.

\section{Problem Formulation}

Consider the case of an investor who has access to $n$ asset classes, each represented by a suitable price index. We define the portfolio optimization problem as an asset allocation problem in which the investor seeks to maximize the return by selecting the appropriate exposure to each asset class while keeping the overall portfolio risk below a predefined   tolerance level.

The notion of risk in the context of financial investments has been widely discussed in  the literature, specifically, the advantages and disadvantages of using different risk metrics.  In our formulation, and in agreement with current best practices, we have chosen the Conditional-Value-at-Risk (CVaR) as a suitable risk metric. Considering that most investors focus on avoiding losses rather than volatility (specially medium- to long-term investors) the CVaR represents a better choice than the standard deviation of returns.  Moreover, the CVaR  (unlike the Value-at-Risk or VaR) has some attractive features, namely, is convex and coherent (i.e., it satisfies the sub-additive condition) \citep{pflug2000some} in the sense of \citet{artzner1999coherent}. 

Let $\textbf{x}\in\mathbb{R}^n$ be the decision vector of weights that specify the asset class allocation   and $\textbf{r}\in\mathbb{R}^n$ the return of each asset class in a given period\footnote{We use boldface font for vectors to distinguish them from scalars.}. The underlying probability distribution of $\textbf{r}$ will be assumed to have a density, which we denote by $\pi(\textbf{r})$. Thus,  the expected return of the portfolio can be expressed as the weighted average of the expected return of each asset class, that is,

\begin{equation}
\label{eqn:weighted_avg}
\begin{aligned}
&     \mathds{E}(\textbf{x}^T\textbf{r}) = \sum^{n}_{i=1} x_i \mathds{E}[r_i] .\\
\end{aligned}
\end{equation}

Let $\alpha \in (0, 1)  \ $ 
be a set level of confidence and $ \Lambda \in\mathbb{R}$ the risk tolerance of the investor.  We can state the optimal portfolio (asset) allocation problem for a  long-only investor as the following optimization problem:

\begin{equation}
\label{eqn:assetalloc_optproblem}
\begin{aligned}
& \underset{\textbf{x} \in \mathbb{R}^n}{\mathrm{maximize}}
& &  \mathds{E}(\textbf{x}^T\textbf{r}) \\
& \mathrm{s.t.}
& &\mathrm{CVaR}_\alpha (\textbf{x}^T\textbf{r})\leq \Lambda \\
&&&  \sum^{n}_{i=1} x_i = 1 \\
&&& \textbf{x}\geq 0.
\end{aligned}
\end{equation}

Recall that the CVaR of a random variable $X$, with a predefined level of confidence $\alpha$, can be expressed as the expected values of the $X$ that exceed the corresponding VaR.  More formally,

\begin{equation}
\label{eqn:cvar}
\begin{aligned}
& \mathrm{CVaR}_\alpha(X) = \frac{1}{1-\alpha}\int_{0}^{1-\alpha} \mathrm{VaR}_\gamma(X) \,d\gamma ,
\end{aligned}
\end{equation}
where $\mathrm{VaR}_\gamma(X)$ denotes the Value-at-Risk of the distribution for a given confidence $\gamma$, defined using the cumulative distribution function  $F_X(x)$   as

\begin{equation}
\label{eqn:var}
\begin{aligned}
& \mathrm{VaR}_\gamma(X) = -\mathrm{inf}\{x\in\mathbb{R}\ |\ F_X(x)>1-\gamma\}.
\end{aligned}
\end{equation}

\begin{subsection}{Discretization and linearization}

In this section, we will explain how the asset allocation problem \eqref{eqn:assetalloc_optproblem}, a nonlinear optimization problem due to the CVaR-based constraint, can be restated as a linear programming problem and why this formulation is useful in practice. 

The foundational concept was established by \cite{rockafellar2000optimization}, who demonstrated that solving the optimization problem \eqref{eqn:assetalloc_optproblem} is equivalent to solving the following optimization problem:

\begin{equation}
\label{eqn:general_continuous_problem}
\begin{aligned}
& \underset{(\textbf{x}, \zeta) \in \mathbb{R}^n \times \mathbb{R} }{\mathrm{minimize}}
& & -\mathds{E}(\textbf{x}^T\textbf{r}) \\
&\mathrm{s.t.} &&  \zeta + (1 - \alpha)^{-1} \int_{\mathbb{R}^n} [-\textbf{x}^T\textbf{r} - \zeta]^+ \pi (\textbf{r}) d\textbf{r} \leq \Lambda \\
&&&  \sum^{n}_{i=1} x_i = 1 \\
&&& \textbf{x}\geq 0.
\end{aligned}
\end{equation}
Here, the dummy variable $\zeta \in \mathbb{R}$ is introduced to serve as a ``threshold for losses''. When $\pi$ has a discrete density $\pi_j$, with $j=1, \ldots, m$ representing the probabilities of occurrence associated to each of the  return vectors or scenarios $\textbf{r}_1, \ldots, \textbf{r}_m$, the problem \eqref{eqn:general_continuous_problem} can be reformulated as:

\begin{equation}
\label{eqn:general_discrete_problem_prev}
\begin{aligned}
& \underset{(\textbf{x}, \zeta) \in \mathbb{R}^n \times \mathbb{R} }{\mathrm{minimize}}
&& -\mathds{E}(\textbf{x}^T\textbf{r}) \\
&\mathrm{s.t.} &&  \zeta + (1 - \alpha)^{-1} \sum_{j=1}^m [-\textbf{x}^T\textbf{r}_j - \zeta]^+ \pi_j \leq \Lambda \\
&&&  \sum^{n}_{i=1} x_i = 1 \\
&&& \textbf{x}\geq 0.
\end{aligned}
\end{equation}
Finally, introducing dummy variables $z_j$, for $j = 1, \ldots, m$, and explicitly writing $\mathds{E}(\textbf{x}^T\textbf{r})$ as $\textbf{x}^T\textbf{R} \bm{\pi}$, where $\textbf{R} \in \mathbb{R}^{n \times m}$ denotes the return variable's range based on the density vector $\bm{\pi}$, the problem in \eqref{eqn:general_discrete_problem_prev} can be restated as the following linear programming problem:

\begin{equation}
\label{eqn:general_discrete_problem}
\begin{aligned}
& \underset{(\textbf{x}, \textbf{z}, \zeta) \in \mathbb{R}^n \times  \mathbb{R}^m \times \mathbb{R}}{\mathrm{maximize}} & & \textbf{x}^T\textbf{R} \bm{\pi} \\
& \mathrm{s.t.} & &\zeta + \frac{1}{1-\alpha} \bm{\pi}^T\textbf{z}  \leq \Lambda \\
&&&\textbf{z} \geq -\textbf{x}^T \textbf{R}  - \zeta \\
&&&  \sum^{n}_{i=1} x_i = 1 \\
&&& \textbf{x}, \textbf{z} \geq 0 .
\end{aligned}
\end{equation}
A comprehensive explanation (including all the formal proofs) regarding the derivation of the equivalent continuous formulation (\ref{eqn:general_continuous_problem}) and its subsequent discretization and linearization, (\ref{eqn:general_discrete_problem_prev}) and (\ref{eqn:general_discrete_problem}), can be found in  \cite{rockafellar2000optimization}. An in-depth analysis of these equivalent formulations--including practical examples, and a discussion of the general problem setting incorporating transaction costs, value constraints, liquidity constraints, and limits on position--are provided in \cite{palmquist1999portfolio}.  

This discretized and linear formulation (\ref{eqn:general_discrete_problem}) has the advantage that it can be handled  with a number of widely available linear optimization solvers. Moreover, the discretization allows us to use sampled data from the relevant probability distribution of $\textbf{r}$ in combination with the appropriate discrete probability density function $\bm{\pi}$.

In this study, as well as in practice, $\textbf{R}$ generally represents a sampled distribution of returns, and the vector $\bm{\pi}$ determines the weights for each of the $m$ sampled return vectors (scenarios). For instance, in the simple case of  random samples without replacement from a set of historical returns,  $\bm{\pi}$ is naturally defined as $\pi_j = 1/m$ for $j \in \{1,...,m\}$. Notice, however, that the weights $\bm{\pi}$ can be modified to adjust the formulation for  a case in which  features (contextual information) are added to the optimization problem. Assume, for example, that we are incorporating a sample of $l$ features $\textbf{F}\in\mathbb{R}^{l\times m}$.  In this case, we redefine $\bm{\pi}$ to reflect the different importance attributed to samples based on the similarity between their corresponding features and those of the current state of the world (economic environment). 

More formally, if $\bm{f_1} $   and  $ \bm{f_2}$  represent normalized\footnote{We say normalized in the sense of transforming the variables into something comparable between them. For simplicity, in our study we used a zero mean normalization (Z-score).} vectors of economic features, we define a distance $d(\cdot, \cdot)$ as

\begin{equation}
\label{eqn:dist}
\begin{aligned}
    d(\bm{f_1}, \bm{f_2}) = \sqrt{(\bm{f_1}- \bm{f_2})^T (\bm{f_1}- \bm{f_2})}.
\end{aligned}
\end{equation}
 Let $\bm{d_f}^{-1}$ be the inverse distance vector of $\textbf{F}$ to a given vector  $\bm{f}$ defined by $[d_f^{-1}]_q = \frac{1}{d(\bm{f}, \bm{f_q})}$, for each $q\in[1,\ldots, m]$, $\bm{f_q}$ being a row of $\bm{F}$. We define the density vector $\bm{\pi}_{\bm{f}}$ as the normalization of the inverse distance vector to $\bm{f}$, that is,

\begin{equation}
\label{eqn:pi}
\begin{aligned}
\bm{\pi}_{\bm{f}} = \frac{\bm{d_f}^{-1}}{\bm{1}^T\bm{d_f}^{-1}}.
\end{aligned}
\end{equation}

For the purpose of this study we will sample $\textbf{R}$ and $\textbf{F}$  based on a suitable  data generator (note that they are not independently sampled, but simultaneously sampled), together with the corresponding weighting scheme (either  $\bm{\pi}$ or the modified density $\bm{\pi}_{\bm{f}}$), and then we will cast the optimization (asset allocation) problem according to the discretized linear framework described in (\ref{eqn:general_discrete_problem}).

\end{subsection}

\section{Synthetic Data Generation}
\label{sec:Method}

In principle, generating random samples from a given  probability density function is a relatively straightforward task. In practice, however, there are two major limitations that prevent finance researchers and practitioners from  relying on such simple  exercise. 

First, and as mentioned before, a financial analyst  has only the benefit of knowing  a  single  path (one   sample outcome) generated by an unknown stochastic process, that is, a multidimensional historical returns time series produced by an unknown data generating process (DGP) (\citet{dgp}).

The second limitation is the non-stationary nature of the stochastic processes underlying  all financial variables. More to the point, financial systems are dynamic and complex, characterized by conditions and mechanisms that vary over time, due to both endogenous effects and external factors (e.g., regulatory changes,  geopolitical events).   Not  surprisingly, the straight reliance on historical data to generate representative scenarios, or, alternatively, attempts based on  conventional (fixed constant) parametric models to generate such scenarios have been disappointing.

 Therefore, given these considerations, our approach consists of using machine learning techniques to generate synthetic data based on recent historical data.  More precisely, the idea is to generate (returns) samples based on a  market-regime aware  generative modeling method known as Conditional Tabular Generative Adversarial Networks (CTGAN).  CTGANs automatically learn and discover patterns in historical data, in an unsupervised mode, to generate realistic synthetic data that mimic  the unknown DGP. Then, we use these generated (synthetic) data to feed the discretized optimization problem described in (\ref{eqn:general_discrete_problem}). 
 
In brief, our goal is to develop a process, that given a historical dataset $\displaystyle \mathcal{D}^{h}$ consisting of $\displaystyle \mathbf{R}^{h}$ asset returns and $\displaystyle \mathbf{F}^{h}$ features (both $\displaystyle m_{h}$ samples), could  train a synthetic data generator (SDG) to create realistic (that is market-regime aware synthetic return datasets) on-demand ($\displaystyle \mathcal{D}^{s}$). Figure \ref{fig:sdg} summarizes visually this concept.

\begin{figure}[H]
\centering
\tbl{The Synthetic Data Generation Schema}

\tikzset{every picture/.style={line width=0.75pt}} 

\begin{tikzpicture}[x=0.75pt,y=0.75pt,yscale=-1,xscale=1]

\draw  [fill={rgb, 255:red, 155; green, 155; blue, 155 }  ,fill opacity=1 ] (187,14.87) -- (233,14.87) -- (233,42.87) -- (187,42.87) -- cycle ;
\draw   (8,8) -- (127,8) -- (127,48) -- (8,48) -- cycle ;
\draw   (294,9) -- (411,9) -- (411,49) -- (294,49) -- cycle ;
\draw    (127,28.87) -- (184,28.38) ;
\draw [shift={(186,28.36)}, rotate = 179.51] [color={rgb, 255:red, 0; green, 0; blue, 0 }  ][line width=0.75]    (10.93,-3.29) .. controls (6.95,-1.4) and (3.31,-0.3) .. (0,0) .. controls (3.31,0.3) and (6.95,1.4) .. (10.93,3.29)   ;
\draw    (233.5,28.61) -- (290.5,28.12) ;
\draw [shift={(292.5,28.11)}, rotate = 179.51] [color={rgb, 255:red, 0; green, 0; blue, 0 }  ][line width=0.75]    (10.93,-3.29) .. controls (6.95,-1.4) and (3.31,-0.3) .. (0,0) .. controls (3.31,0.3) and (6.95,1.4) .. (10.93,3.29)   ;

\draw (197,20.4) node [anchor=north west][inner sep=0.75pt]    {$\mathtt{SDG}$};
\draw (15,16.4) node [anchor=north west][inner sep=0.75pt]    {$\mathcal{D}^{h} \ =\left[\mathbf{R}^{h} \ \mathbf{F}^{h}\right]$};
\draw (301,17.4) node [anchor=north west][inner sep=0.75pt]    {$\mathcal{D}^{s} \ =\left[\mathbf{R}^{s} \ \mathbf{F}^{s}\right]$};
\draw (143,17) node [anchor=north west][inner sep=0.75pt]  [font=\scriptsize] [align=left] {train};
\draw (235,15.87) node [anchor=north west][inner sep=0.75pt]  [font=\scriptsize] [align=left] {generate};

\end{tikzpicture}

\label{fig:sdg}

\end{figure}

\subsection{Conditional Tabular Generative Adversarial Networks (CTGAN)}

Recent advances in machine learning and neural networks, specifically, the development of Generative Adversarial Networks (GAN) that can mix continuous and discrete (tabular) data to generate regime-aware samples, are particularly useful in financial engineering applications.  A good example is the method proposed by  \citet{leixu2019}.  These authors   introduced a neural network architecture for Conditional Tabular Generative Adversarial Networks (CTGAN) to generate synthetic data.   This approach  presents several advantages, namely, it can create  a realistic synthetic data generating process  that can capture, in our case,  the complex relationships between asset returns and features, while being sensitive to the existence of different market regimes.\\

In general terms, the architecture of a CTGAN differs from that of a standard GAN  in several ways:

\begin{itemize}

\item CTGAN models the dataset as a conditional process, where the continuous variables are defined by a conditional distribution dependent on the discrete variables, and each combination of discrete variables defines a state that determines the uni and multivariate distribution of the continuous variables.

\item To avoid the problem of class imbalances in the training process, CTGAN introduces the notions of conditional generator and a trainning-by-sampling process. The conditional generator decomposes the probability distribution of a sample as the aggregation of the conditional distributions given all the possible discrete values for a selected variable.
Given this decomposition, a conditional generator can be trained considering each specific discrete state, allowing the possibility of a training-by-sampling process that can select states evenly for the conditional generator and avoid poor representation of low-frequency states.

\item CTGAN improves the normalization of the continuous columns employing mode-specific normalization. For each continuous variable, the model uses Variational Gaussian Mixture models to identify the different modes of its univariate distribution and decompose each sample using the normalized value based on the most likely mode and a one hot-vector defined by the mode used. This process improves the suitability of the dataset for training, converting it into a bounded vector representation easier to process by the network.  

\end{itemize}

\subsection{A Modified CTGAN-plus-features method}

To enhance the capacity of generating state-aware synthetic data (scenarios) based on the CTGAN architecture, we use an unsupervised method to generate discrete market regimes or states. Our approach is based on identifying clusters of samples exhibiting similar characteristics in terms of asset returns and  features, and we   finally  use the cluster identifier as the state-defining variable employed by the CTGAN model.

A full discussion of how to generate a market regime (or state)   aware identification model goes beyond the scope of this study.  Suffice to say that in this case  we relied on  well-known methods from the machine learning literature for dimensionality reduction such as t-NSE (short for  t-distributed Stochastic Neighbor Embedding)     and density-based clustering   such as  HDBSCAN    (short for  High Density-Based Spatial Clustering of Applications with Noise.)          \citep{campello2013density}. Additionally, to reduce the noise generated by trivially-correlated assets (like the S\&P 500 and the Nasdaq 100, for example), we first decompose the asset returns based on their principal components using a  PCA technique (where the number of dimensions is equal to the number of asset classes).\\

In summary, the synthetic data generating process, which is  described schematically in Figure \ref{fig:sdg2}, consists of the following steps:

\begin{enumerate}
\item Start with an historical dataset $\displaystyle \mathcal{D}^{h}$ consisting of  $\displaystyle \mathbf{R}^{h}$ asset returns and $\displaystyle \mathbf{F}^{h}$ features (both $\displaystyle m_{h}$ samples from the same periods).
\item The dataset is orthogonalyzed in all its principal components using PCA to avoid forcing the model to estimate the dependency of highly correlated assets such as equity indexes with mayor overlaps.   The eigenvectors are stored to reverse the projection on the synthetically generated dataset.
\item Generate a discrete vector \textbf{C} assigning a cluster identifier  to each sample. The process to generate the clusters consists of two steps:
\begin{enumerate}
\item Reduce the dimensionality of the dataset $\displaystyle \mathcal{D}^{h}$ from $\displaystyle m_{h}$ to 2 using t-SNE.
\item Apply HDBSCAN on the 2-dimensional projection of $\displaystyle \mathcal{D}^{h}$.
\end{enumerate}
\item Train a CTGAN  using as continuous variables the PCA-transformed dataset ($\displaystyle \mathcal{D}_{pca}^{h}$) and the vector \textbf{C} as an extra discrete column of the dataset.
\item Generate $\displaystyle m_{s}$ synthetic samples using the trained CTGAN ($\displaystyle \mathcal{D}_{pca}^{s}$).
\item Reverse the projection from the PCA space to its original space in the synthetic dataset $\displaystyle \mathcal{D}_{pca}^{s}$ using the stored eigenvectors, obtaining a new synthetic dataset $\displaystyle \mathcal{D}^{s}$ of $\displaystyle m_{s}$ samples.
\end{enumerate}

\begin{figure}[H]
\centering
\tbl{The Modified CTGAN-plus-Features Data Generating Process}

\tikzset{every picture/.style={line width=0.75pt}} 

\begin{tikzpicture}[x=0.75pt,y=0.75pt,yscale=-1,xscale=1]

\draw   (5.13,11) -- (240.5,11) -- (240.5,55.54) -- (5.13,55.54) -- cycle ;
\draw   (4,104.39) -- (239.37,104.39) -- (239.37,133.12) -- (4,133.12) -- cycle ;

\draw    (120,55.54) -- (120,98) ;
\draw [shift={(120,100)}, rotate = 269.99] [color={rgb, 255:red, 0; green, 0; blue, 0 }  ][line width=0.75]    (10.93,-3.29) .. controls (6.95,-1.4) and (3.31,-0.3) .. (0,0) .. controls (3.31,0.3) and (6.95,1.4) .. (10.93,3.29)   ;
\draw   (370.32,105.11) -- (472.8,105.11) -- (472.8,133.84) -- (370.32,133.84) -- cycle ;
\draw    (240.5,119.47) -- (365,119.99) ;
\draw [shift={(367,120)}, rotate = 180.24] [color={rgb, 255:red, 0; green, 0; blue, 0 }  ][line width=0.75]    (10.93,-3.29) .. controls (6.95,-1.4) and (3.31,-0.3) .. (0,0) .. controls (3.31,0.3) and (6.95,1.4) .. (10.93,3.29)   ;
\draw   (146,184) -- (473,184) -- (473,215.73) -- (146,215.73) -- cycle ;
\draw    (434.97,133.84) -- (435,179) ;
\draw [shift={(435,181)}, rotate = 269.96] [color={rgb, 255:red, 0; green, 0; blue, 0 }  ][line width=0.75]    (10.93,-3.29) .. controls (6.95,-1.4) and (3.31,-0.3) .. (0,0) .. controls (3.31,0.3) and (6.95,1.4) .. (10.93,3.29)   ;
\draw   (202.01,243.75) -- (350.67,243.75) -- (350.67,273.2) -- (202.01,273.2) -- cycle ;
\draw    (278,216) -- (278,239) ;
\draw [shift={(278,241)}, rotate = 270] [color={rgb, 255:red, 0; green, 0; blue, 0 }  ][line width=0.75]    (10.93,-3.29) .. controls (6.95,-1.4) and (3.31,-0.3) .. (0,0) .. controls (3.31,0.3) and (6.95,1.4) .. (10.93,3.29)   ;
\draw   (181.74,302.94) -- (453,302.94) -- (453,335) -- (181.74,335) -- cycle ;
\draw    (188.97,132.84) -- (189,178) ;
\draw [shift={(189,180)}, rotate = 269.96] [color={rgb, 255:red, 0; green, 0; blue, 0 }  ][line width=0.75]    (10.93,-3.29) .. controls (6.95,-1.4) and (3.31,-0.3) .. (0,0) .. controls (3.31,0.3) and (6.95,1.4) .. (10.93,3.29)   ;
\draw    (278,274) -- (278,297) ;
\draw [shift={(278,299)}, rotate = 270] [color={rgb, 255:red, 0; green, 0; blue, 0 }  ][line width=0.75]    (10.93,-3.29) .. controls (6.95,-1.4) and (3.31,-0.3) .. (0,0) .. controls (3.31,0.3) and (6.95,1.4) .. (10.93,3.29)   ;

\draw (384.12,107.01) node [anchor=north west][inner sep=0.75pt]    {$\mathbf{C}^{h} \in \ \mathbb{N}_{k}^{m_{h}}$};
\draw (152.09,188.48) node [anchor=north west][inner sep=0.75pt]    {$\mathtt{CTGAN}\left( continous=\mathcal{D}_{pca}^{h} \ ,\ discrete=\mathbf{C}^{h}\right)$};
\draw (222.65,247.23) node [anchor=north west][inner sep=0.75pt]    {$\mathcal{D}_{pca}^{s} \ \in \mathbb{R}^{n+l,\ m_{s}} \ $};
\draw (34.12,105.65) node [anchor=north west][inner sep=0.75pt]    {$\mathcal{D}_{pca}^{h} \coloneqq \mathtt{PCA}\left(\left[\mathbf{R}^{h} \ \mathbf{F}^{h}\right]\right)$};
\draw (124.86,65.24) node [anchor=north west][inner sep=0.75pt]  [font=\scriptsize] [align=left] {Normalize and \\orthogonalize via PCA};
\draw (258.84,104.15) node [anchor=north west][inner sep=0.75pt]  [font=\scriptsize] [align=left] {Clustering Process};
\draw (191.34,153.29) node [anchor=north west][inner sep=0.75pt]  [font=\scriptsize] [align=left] {Train CTGAN using C\textsuperscript{h} as the discrete variable};
\draw (279.46,219.01) node [anchor=north west][inner sep=0.75pt]  [font=\scriptsize] [align=left] {Generate synthetic samples};
\draw (280,277) node [anchor=north west][inner sep=0.75pt]  [font=\scriptsize] [align=left] {Reverse PCA using original eigenvectors};
\draw (26.52,32.01) node [anchor=north west][inner sep=0.75pt]    {$\mathcal{D}^{h} \ =\left[\mathbf{R}^{h} \ \mathbf{F}^{h}\right] \ \in \ \mathbb{R}^{n+l,\ m_{h}}$};
\draw (51.37,12.92) node [anchor=north west][inner sep=0.75pt]  [font=\scriptsize] [align=left] {Start with an historical dataset };
\draw (197,307.4) node [anchor=north west][inner sep=0.75pt]    {$\mathtt{PCA}^{-1}\left(\mathcal{D}_{pca}^{s}\right) \ =\ \left[\mathbf{R}^{s} \ \mathbf{F}^{s}\right] \ =\ \mathcal{D}^{s}$};

\end{tikzpicture}

\label{fig:sdg2}
\end{figure}

\section{     Example of Application}

The following example will help  to assess the merits of our approach  vis-à-vis other alternative asset allocation schemes.  Consider the case of an investor who has access to ten asset classes (a diverse assortment of stocks, bonds and commodities) based on the indices described in Table \ref{indices}. We further assume that the investor has a medium- to long-term horizon and that he/she will be rebalancing his/her portfolio (recalculating the asset allocation weights) once a year, which for simplicity we assume that is done at the beginning of the calendar year (January). We consider the period January 2003-June 2022, a time span for which we have gathered daily returns  data corresponding to all the indices listed in Table \ref{indices}.  Finally, we assume that the investor will rely on a 5-year lookback period to, first, generate synthetic returns data (via the Modified CTGAN approach outlined in the previous section), and then, would rely on the linear optimization framework described in (\ref{eqn:general_discrete_problem}) to determine the asset allocation weights.

\renewcommand{\arraystretch}{1.3}
\begin{table}[H]
\tbl{Indices Employed in the Asset Allocation Example}
{    \begin{tabular}{ccc}\hline
        \textbf{Asset Class} & \textbf{Bloomberg Ticker} & \textbf{Name}  \\ \hline
        US Equities & SPX & S\&P 500 Index \\
        US Equities Tech & NDX &  Nasdaq 100 Index \\
        Global Equities & MXWO & Total Stock Market Index \\
        EM Equities & MXEF & Emerging Markets Stock Index \\ \hline 
        High Yield & IBOXHY & High Yield Bonds Index \\
        Investment Grade & IBOXIG & Liquid Investment Grade Index  \\
        EM Debt & JPEIDIVR &  Emerging Markets Bond Index \\ \hline 
        Commodities & BCOMTR & Bloomberg Commodity Index \\  \hline
        Long-term Treasuries & I01303US & Long-Term Treasury Index \\
        Short-term Treasuries & LT01TRUU & Short-Term Treasury Index \\  \hline
    \end{tabular}
}
\label{indices}
\end{table}

\subsection{Feature selection}

As mentioned before, incorporating features to an optimization problem can greatly improve the out-of-sample performance of the solutions.  Financial markets offer a huge number of options for contextual information.  The list is long and includes macroeconomic indicators, such as GDP, consumer confidence indices, or retail sales volume.  Since our intention is to incorporate an indicator that could describe the state of the economy at several specific times, we argue that the Treasury yield curve (or more precisely, the interest rates corresponding to different maturities) is a suitable choice for several reasons.  First, the yield curve is very dynamic as it quickly reflects changes in market conditions, as opposed to other indicators which are calculated on a monthly or weekly basis and take  more time to adjust.  Second, its computation is “error-free” in the sense that is not subject to ambiguous interpretations or subjective definitions such as the unemployment rate or construction spending.  And third, it summarizes the overall macroeconomic environment—not just one aspect of it—while offering some implicit predictions regarding the direction the economy is moving.  In fact, both the empirical evidence and much of the academic literature, support the view that the yield curve (also known as the term structure of interest rates) is a useful tool for estimating the likelihood of a future recession, pricing financial assets, guiding monetary policy, and forecasting economic growth.  A discussion of the yield curve with reference to its information content is beyond the scope of this paper.  However, a number of studies have covered this issue extensively (e.g., \citet{kumar2021relationship, bauer2018information, evgenidis2020yield, estrella2006yield}).  For the purpose of this example we use the U.S. yield curve tenors specified in Table \ref{features}.  In other words, we use eight features, and each feature corresponds to the  interest rate associated with a different maturity.

\begin{table}[H]
\tbl{Features (Index Returns) Used in the Asset Allocation Example}
{    \begin{tabular}{c@{\hspace{5em}}c@{\hspace{5em}}}\hline
        Bloomberg Ticker & Maturity  \\ \hline
        FDTR & 0 Months (Fed funds rate) \\
        I02503M &  3 Months \\
        I02506M &  6 Months\\
        I02501Y &  1 Year \\
        I02502Y &  2 Years \\
        I02505Y &  5 Years \\
        I02510Y &  10 Years \\
        I02530Y &  30 Years \\

    \end{tabular}}
\label{features}
\end{table}

\subsection{Synthetic Data Generation Process (SDGP) Validation }

Given the paramount importance played by the synthetic data generation process (SDGP) in our approach, it makes sense, before solving any optimization problem, to investigate  whether the CTGAN model actually generates suitable scenarios (or data samples).  In other words, to explore if the quality of the SDGP is appropriate to mimic the unknown stochastic process behind the historical data.  Although the inner structure of the actual stochastic process is unknown, one can always compare the similarity between the input and output distributions.  In short, we can compare if  their single and joint multivariate distributions are similar, and that the synthetic samples are not an exact copy of the (original) training samples.  To perform this comparison, we trained the CTGAN using  historical data from the 2017-2022 period (5 years).

\begin{figure}[H]
\centering
\resizebox*{13cm}{!}{\includegraphics{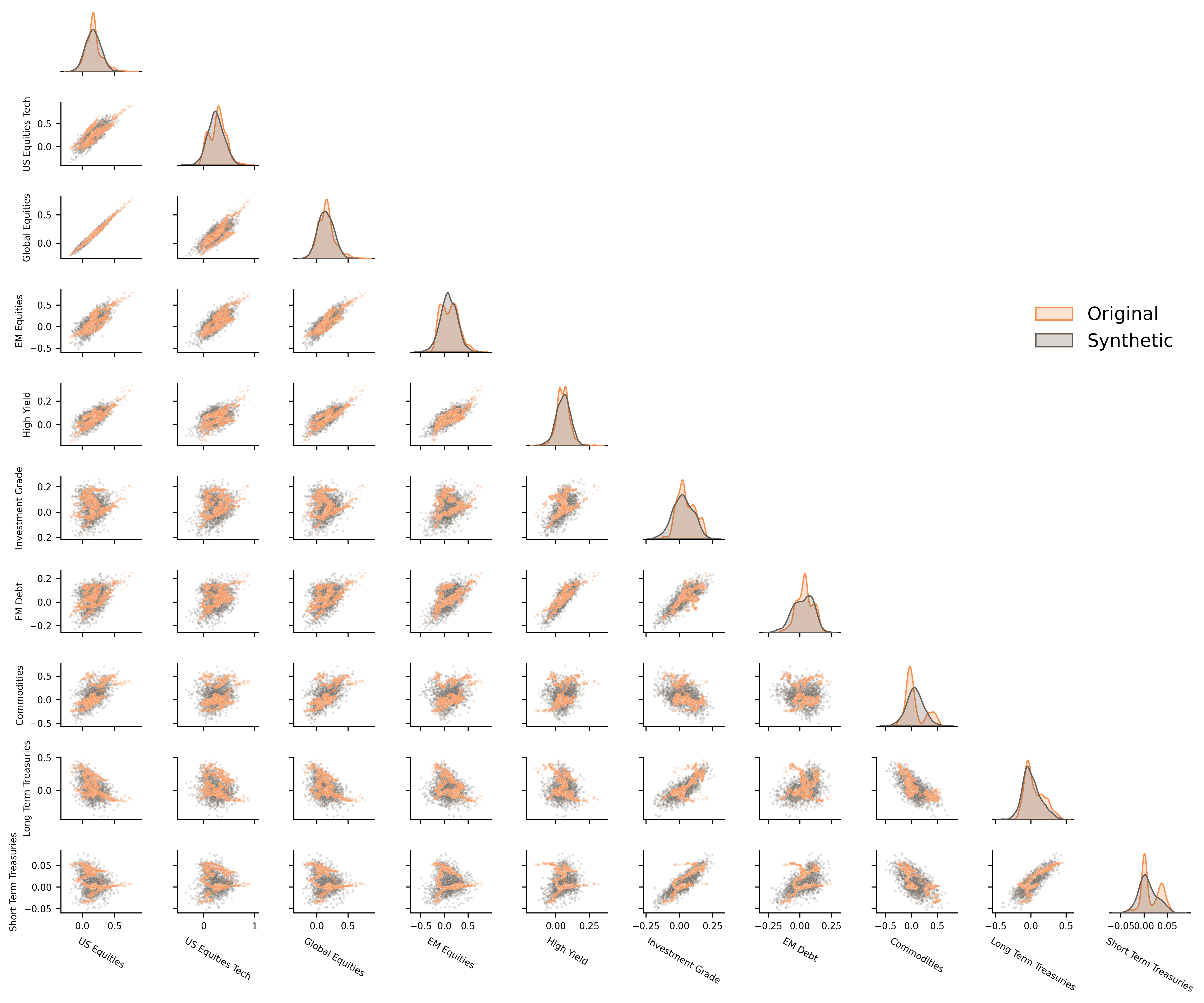}}
\caption{Pair-Plot comparison of synthetic versus original data, annual returns.}
\label{fig:pairasset}
\end{figure}

\begin{figure}[H] 
\centering
\resizebox*{13cm}{!}{\includegraphics{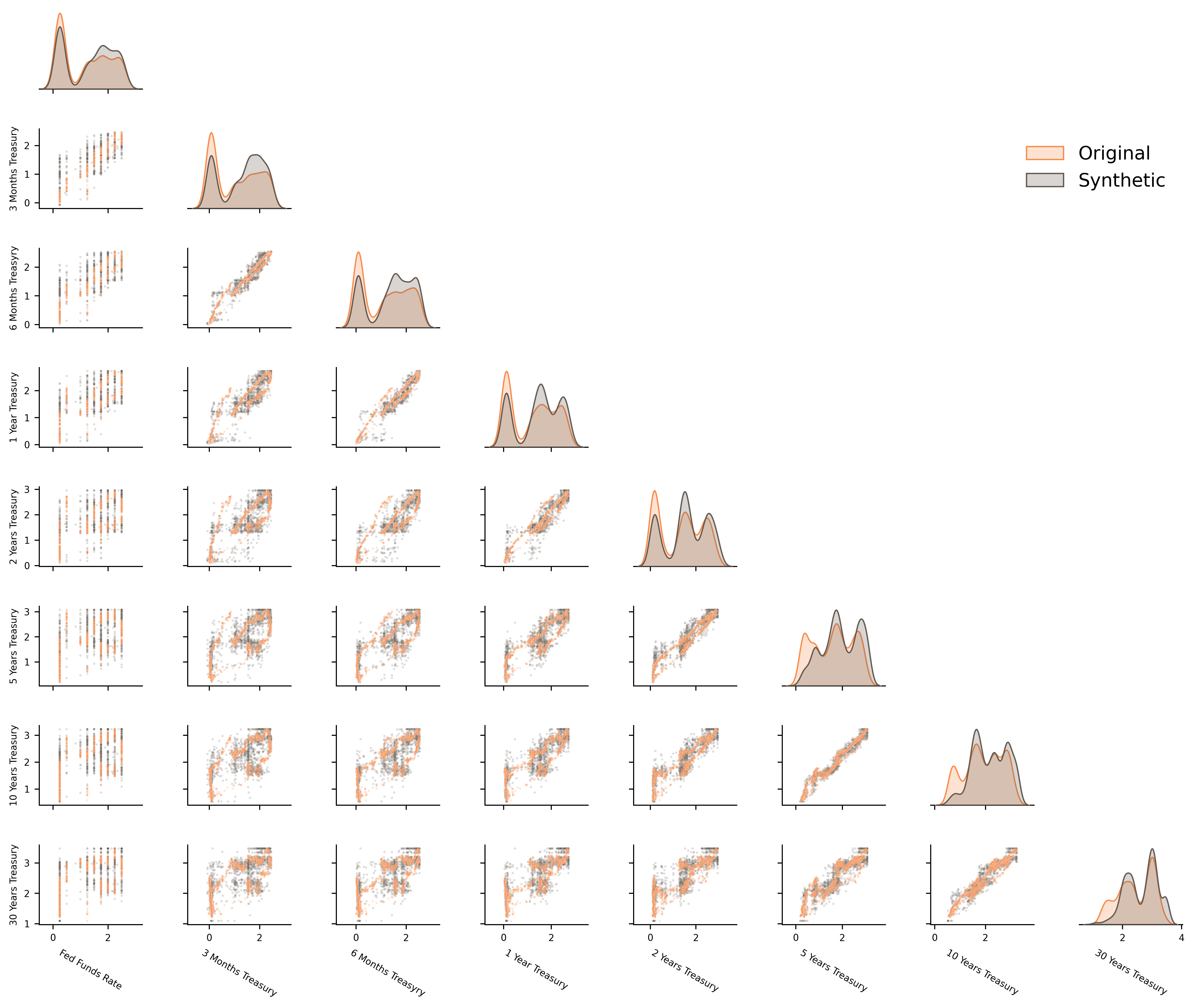}}
\caption{Pair-Plot comparison of  synthetic versus original data features, annual yields.}
\label{fig:pairfeature}
\end{figure}

Figures \ref{fig:pairasset} and \ref{fig:pairfeature} both hint that the synthetic data actually display the same characteristics of the original data.  However, and notwithstanding the compelling visual evidence, it is possible to make a more quantitative assessment to validate the SDGP. To this end, we can perform two comparisons. First, we can compare for each variable (e.g., U.S. equities returns) the corresponding marginal distribution based on the original and synthetic data to see if they are  indeed similar.  And second, for each pair of variables, we can compare the corresponding joint distributions.

Table \ref{univariate-dist} reports the results of the  Kolmogorov-Smirnov test (KS-test) \citep{kolmogorov},  which seeks to determine whether both samples (original and synthetic) come from the same distribution.  The null hypothesis (e.g., that both samples come from the same distribution) cannot be rejected.  Notice that the table reports the complement score, that is, a value of 1 refers to two identical distributions while 0 signals two different distributions.  The average value is 0.87, suggesting that in all cases, both the original and synthetic distributions, are very similar in nature.

    \begin{table}[H]
    \tbl{  Kolmogorov-Smirnov Test:  Comparison Between Original and Synthetic Returns and Interest Rates Distributions }
    {    \begin{tabular}{cccccc}\hline
    Variable                        &   KS-test Score  & Variable                      &   KS-test Score \\ \hline
    US Equities                     &   91.89\%       & Fed Funds Rate          &   89.21\%      \\
    US Equities Tech         &   86.30\%       & 3 Months Treasury         &   82.85\%      \\
    Global Equities             &   94.52\%       & 6 Months Treasury        &   82.58\%      \\
    EM equities             &   92.66\%       & 1 Year Treasury        &   84.44\%      \\
    High Yield                   &   93.53\%       & 2 Years Treasury        &   86.41\%      \\
    Investment Grade                  &   85.87\%       & 5 Years Treasury       &   84.61\%      \\
    EM Debt                 &   86.47\%        & 10 Years Treasury       &   85.87\%      \\
    Commodities              &   76.61\%      & 30 Years Treasury      &   85.21\%      \\
    Long-term Treasuries   &   88.11\%       &                               &                 \\
    Short-term Treasuries  &   80.55\%       &                               &                 \\
        \end{tabular}}
    \label{univariate-dist}
    
    \end{table}

\begin{figure}[H]
    \centering
    \resizebox*{10cm}{!}{\includegraphics{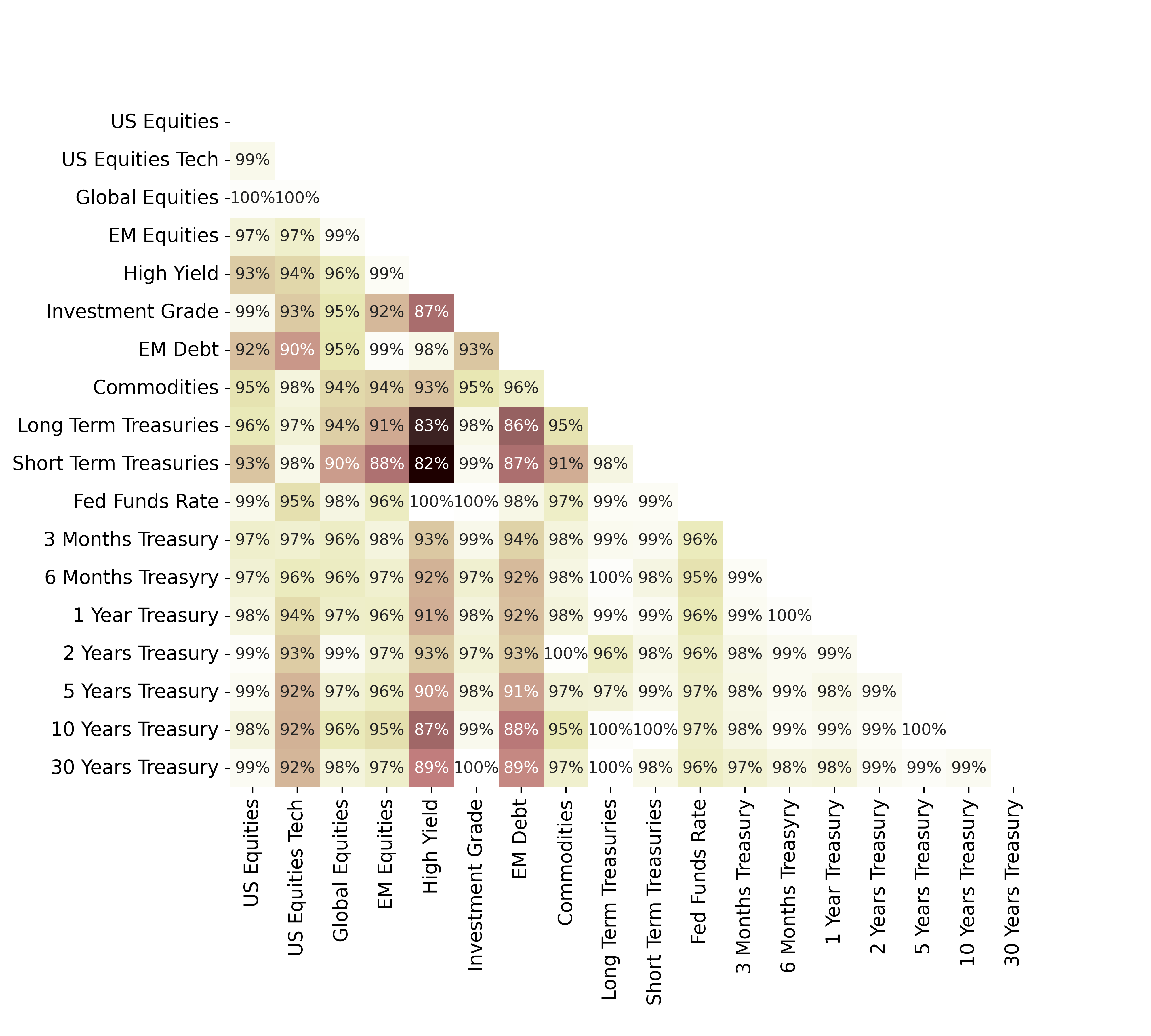}}
    \caption{Correlation similarity comparison between the correlation matrices of the original and the synthetic data.}
     \label{fig:dist}
    \end{figure}

  In order to verify that the synthetic samples preserve the relationship that existed between the variables in the original data, we compared the joint distributions based on the original and synthetic datasets. To this end, we compared the degree of similarity of the correlation  matrices determined by each sample. Specifically, for any two variables, say, for example, US Equities and Commodities, we would expect the correlation between them to be similar in both, the original and synthetic datasets.  Figure (\ref{fig:dist}) shows, for all possible paired-comparisons, the value of a correlation similarity index,  Such index is defined as 1 minus the absolute value of the difference between both (original and synthetic data) correlations. A value of 1 indicates identical values; a value of 0 indicates a maximum discrepancy.  The values shown in Figure (\ref{fig:dist}) (the lowest is 0.83) evidence a high level of agreement.

\begin{figure}[H] 
\centering
\resizebox*{13cm}{!}{\includegraphics{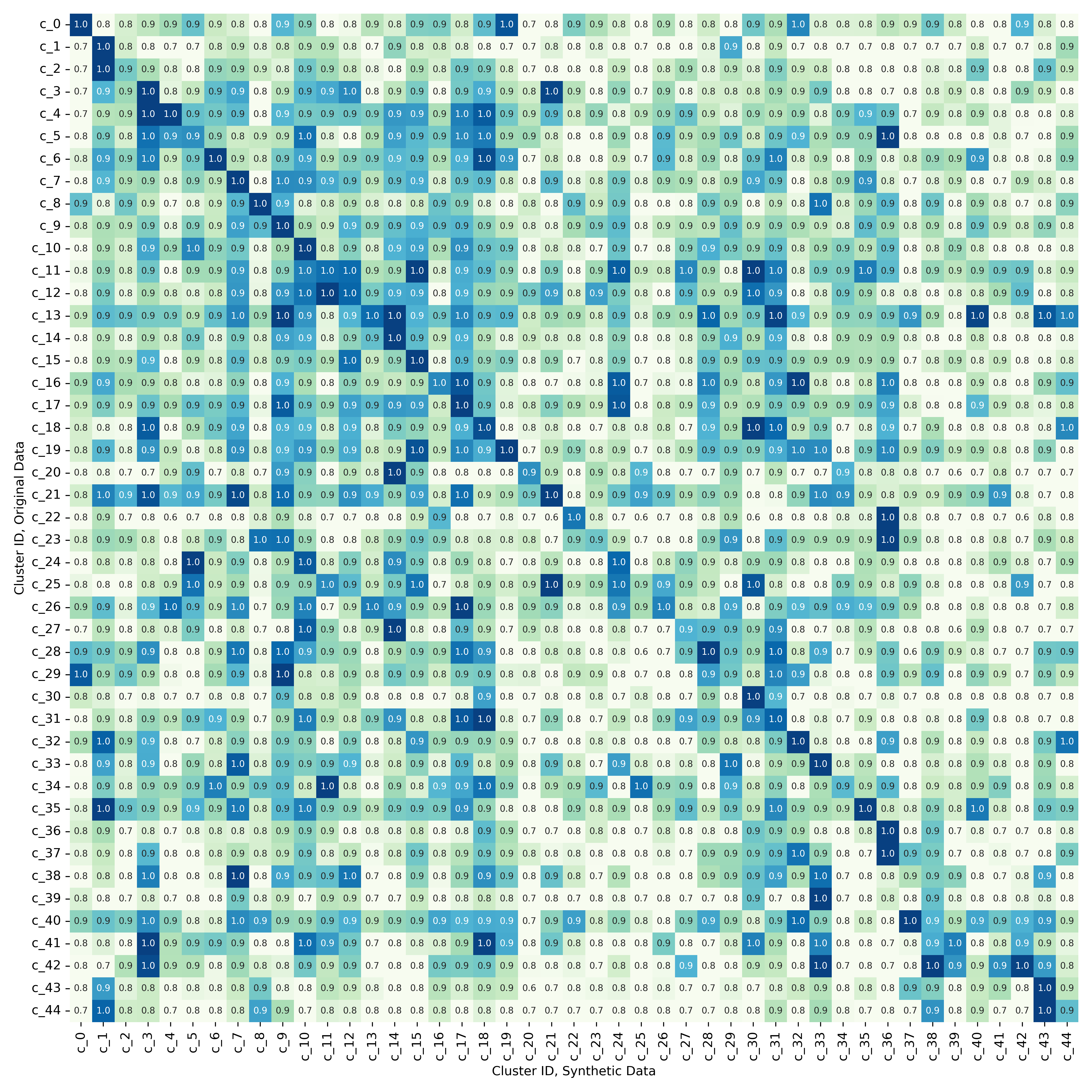}}
\caption{Pair-Plot comparison of  synthetic versus original data average KS-Test across all dimensions, divided by cluster. Values are scaled by the maximum KS-test score of each row}
\label{fig:pairfeature2}
\end{figure}

A more nuanced comparison between the characteristics of the original (historical) dataset and the synthetic dataset can be accomplished by looking at the clusters. In other words, the different market regimes identified during the data generation process.  This comparison can be carried out in two steps.  First, we computed the correlation between the distribution of data points across clusters in the original dataset and their counterparts in the synthetic dataset. The number of synthetic samples drawn from each cluster followed a distribution that closely mirrors the distribution of clusters identified in the original dataset (44 clusters in total), having a correlation of 97.2\%. This high degree of agreement can be attributed to the CTGAN's training process, wherein the probability distribution for the conditional variables are explicitly learned, facilitating an accurate replication of the original dataset's structural characteristics. And second, we refined the KS-test, partitioning both the original and synthetic datasets based on their  respective clusters. This allowed us to compare the similarities between samples where the original and synthetic data originated from the same cluster versus those from different clusters. The results of this exercise, displayed in  Figure \ref{fig:pairfeature2}, reveal that the synthetic data conditioned on the same cluster as the original data typically yielded the highest KS-test scores compared to data generated from other clusters. This finding provides further evidence of the effectiveness  of the cluster-based approach to produce synthetic data that  replicates not only the broad characteristics of the original dataset but also all the key elements of all the different market regimes.

In conclusion, based on the previous results we can state with confidence that the CTGAN does create data samples   congruent with the original dataset, effectively preserving both marginal and joint distributions. Furthermore, our results highlight a tangible improvement in the quality of data generation attributable to the incorporation of the clustering process. Having validated the SDGP,  the next step is to assess the merits of the optimization approach itself.

\subsection{Testing  strategy }

In order to better assess the performance of our approach, i.e., (Modified) CTGAN with features, which we denote as GwF, we compare it with four additional asset allocation strategies, as indicated below.  In short, we test five strategies, namely: 

\begin{enumerate}
  \item[(i)] CTGAN without features (Gw/oF)
  \item[(ii)] CTGAN with features (GwF)
  \item[(iii)] Historical data without features (Hw/oF)
  \item[(iv)] Historical data with features (HwF)
  \item[(v)] Equal Weights (EW)
\end{enumerate}

The historical-data strategies, unlike the CTGAN-based strategies, are based on direct sampling from historical data. We also utilize the Equal-Weight (EW) strategy, known as the 1/N strategy, which assigns equal weights to all asset classes. This approach is chosen precisely because it does not depend on any predetermined risk constraint or measure, nor does it rely on historical data. Its effectiveness is not contingent on the assumptions required by other strategies that use measures like CVaR to bound risk. Despite its simplicity, this seemingly naive strategy has generally performed surprisingly well, often outperforming many variations of Mean-Variance (MV) strategies. A comprehensive evaluation of the EW strategy's performance can be found in the work of \citet{demiguel2009optimal}, which underscores its utility as a useful benchmark. Indeed, we contend that any strategy failing to outperform the EW strategy likely has little to offer and is unlikely to be of practical relevance.

\begin{figure}[H]
\centering
\tbl{ Sequence of 5-year Overlapping Windows }

\tikzset{every picture/.style={line width=0.75pt}} 

\begin{tikzpicture}[x=0.75pt,y=0.75pt,yscale=-1,xscale=1]

\draw [color={rgb, 255:red, 74; green, 74; blue, 74 }  ,draw opacity=1 ]   (53.17,150) -- (348.2,150.5) ;
\draw [shift={(351.2,150.5)}, rotate = 180.1] [fill={rgb, 255:red, 74; green, 74; blue, 74 }  ,fill opacity=1 ][line width=0.08]  [draw opacity=0] (8.93,-4.29) -- (0,0) -- (8.93,4.29) -- cycle    ;
\draw [shift={(50.17,150)}, rotate = 0.1] [fill={rgb, 255:red, 74; green, 74; blue, 74 }  ,fill opacity=1 ][line width=0.08]  [draw opacity=0] (8.93,-4.29) -- (0,0) -- (8.93,4.29) -- cycle    ;
\draw [line width=1.5]    (38.8,170.5) -- (558,170.45) ;
\draw [shift={(562,170.45)}, rotate = 180] [fill={rgb, 255:red, 0; green, 0; blue, 0 }  ][line width=0.08]  [draw opacity=0] (13.4,-6.43) -- (0,0) -- (13.4,6.44) -- (8.9,0) -- cycle    ;
\draw [shift={(34.8,170.5)}, rotate = 360] [fill={rgb, 255:red, 0; green, 0; blue, 0 }  ][line width=0.08]  [draw opacity=0] (13.4,-6.43) -- (0,0) -- (13.4,6.44) -- (8.9,0) -- cycle    ;
\draw [line width=1.5]    (50.36,160.05) -- (50.18,180.23) ;
\draw [line width=1.5]    (110.9,160.05) -- (110.72,180.23) ;
\draw [line width=1.5]    (170.63,160.11) -- (170.45,180.29) ;
\draw [line width=1.5]    (230.83,160.05) -- (230.65,180.23) ;
\draw [line width=1.5]    (290.9,160.31) -- (290.72,180.49) ;
\draw [line width=1.5]    (350.7,160.78) -- (350.52,180.96) ;
\draw [line width=1.5]    (411.23,160.05) -- (411.05,180.23) ;
\draw [line width=1.5]    (471.23,160.45) -- (471.05,180.63) ;
\draw [color={rgb, 255:red, 74; green, 74; blue, 74 }  ,draw opacity=1 ]   (113.57,140) -- (408.6,140.5) ;
\draw [shift={(411.6,140.5)}, rotate = 180.1] [fill={rgb, 255:red, 74; green, 74; blue, 74 }  ,fill opacity=1 ][line width=0.08]  [draw opacity=0] (8.93,-4.29) -- (0,0) -- (8.93,4.29) -- cycle    ;
\draw [shift={(110.57,140)}, rotate = 0.1] [fill={rgb, 255:red, 74; green, 74; blue, 74 }  ,fill opacity=1 ][line width=0.08]  [draw opacity=0] (8.93,-4.29) -- (0,0) -- (8.93,4.29) -- cycle    ;
\draw [color={rgb, 255:red, 74; green, 74; blue, 74 }  ,draw opacity=1 ]   (173.97,130.4) -- (469,130.9) ;
\draw [shift={(472,130.9)}, rotate = 180.1] [fill={rgb, 255:red, 74; green, 74; blue, 74 }  ,fill opacity=1 ][line width=0.08]  [draw opacity=0] (8.93,-4.29) -- (0,0) -- (8.93,4.29) -- cycle    ;
\draw [shift={(170.97,130.4)}, rotate = 0.1] [fill={rgb, 255:red, 74; green, 74; blue, 74 }  ,fill opacity=1 ][line width=0.08]  [draw opacity=0] (8.93,-4.29) -- (0,0) -- (8.93,4.29) -- cycle    ;
\draw [color={rgb, 255:red, 74; green, 74; blue, 74 }  ,draw opacity=1 ]   (232.77,120) -- (527.8,120.5) ;
\draw [shift={(530.8,120.5)}, rotate = 180.1] [fill={rgb, 255:red, 74; green, 74; blue, 74 }  ,fill opacity=1 ][line width=0.08]  [draw opacity=0] (8.93,-4.29) -- (0,0) -- (8.93,4.29) -- cycle    ;
\draw [shift={(229.77,120)}, rotate = 0.1] [fill={rgb, 255:red, 74; green, 74; blue, 74 }  ,fill opacity=1 ][line width=0.08]  [draw opacity=0] (8.93,-4.29) -- (0,0) -- (8.93,4.29) -- cycle    ;
\draw [line width=1.5]    (531.23,159.25) -- (531.05,179.43) ;
\draw [color={rgb, 255:red, 155; green, 155; blue, 155 }  ,draw opacity=1 ][line width=0.75]    (60.2,164.87) -- (60.07,175.23) ;
\draw [color={rgb, 255:red, 155; green, 155; blue, 155 }  ,draw opacity=1 ]   (55.4,165) -- (55.27,175.37) ;
\draw [color={rgb, 255:red, 155; green, 155; blue, 155 }  ,draw opacity=1 ][line width=0.75]    (80.2,164.87) -- (80.07,175.23) ;
\draw [color={rgb, 255:red, 155; green, 155; blue, 155 }  ,draw opacity=1 ]   (75.4,165) -- (75.27,175.37) ;
\draw [color={rgb, 255:red, 155; green, 155; blue, 155 }  ,draw opacity=1 ][line width=0.75]    (70.07,164.73) -- (69.93,175.1) ;
\draw [color={rgb, 255:red, 155; green, 155; blue, 155 }  ,draw opacity=1 ]   (65.27,164.87) -- (65.13,175.23) ;
\draw [color={rgb, 255:red, 155; green, 155; blue, 155 }  ,draw opacity=1 ][line width=0.75]    (90.2,164.73) -- (90.07,175.1) ;
\draw [color={rgb, 255:red, 155; green, 155; blue, 155 }  ,draw opacity=1 ]   (85.4,164.87) -- (85.27,175.23) ;
\draw [color={rgb, 255:red, 155; green, 155; blue, 155 }  ,draw opacity=1 ]   (105.4,164.87) -- (105.27,175.23) ;
\draw [color={rgb, 255:red, 155; green, 155; blue, 155 }  ,draw opacity=1 ][line width=0.75]    (100.07,164.6) -- (99.93,174.97) ;
\draw [color={rgb, 255:red, 155; green, 155; blue, 155 }  ,draw opacity=1 ]   (95.27,164.73) -- (95.13,175.1) ;
\draw [color={rgb, 255:red, 155; green, 155; blue, 155 }  ,draw opacity=1 ][line width=0.75]    (119.8,164.73) -- (119.67,175.1) ;
\draw [color={rgb, 255:red, 155; green, 155; blue, 155 }  ,draw opacity=1 ]   (115,164.87) -- (114.87,175.23) ;
\draw [color={rgb, 255:red, 155; green, 155; blue, 155 }  ,draw opacity=1 ][line width=0.75]    (139.8,164.73) -- (139.67,175.1) ;
\draw [color={rgb, 255:red, 155; green, 155; blue, 155 }  ,draw opacity=1 ]   (135,164.87) -- (134.87,175.23) ;
\draw [color={rgb, 255:red, 155; green, 155; blue, 155 }  ,draw opacity=1 ][line width=0.75]    (129.67,164.6) -- (129.53,174.97) ;
\draw [color={rgb, 255:red, 155; green, 155; blue, 155 }  ,draw opacity=1 ]   (124.87,164.73) -- (124.73,175.1) ;
\draw [color={rgb, 255:red, 155; green, 155; blue, 155 }  ,draw opacity=1 ][line width=0.75]    (149.8,164.6) -- (149.67,174.97) ;
\draw [color={rgb, 255:red, 155; green, 155; blue, 155 }  ,draw opacity=1 ]   (145,164.73) -- (144.87,175.1) ;
\draw [color={rgb, 255:red, 155; green, 155; blue, 155 }  ,draw opacity=1 ]   (165,164.73) -- (164.87,175.1) ;
\draw [color={rgb, 255:red, 155; green, 155; blue, 155 }  ,draw opacity=1 ][line width=0.75]    (159.67,164.47) -- (159.53,174.83) ;
\draw [color={rgb, 255:red, 155; green, 155; blue, 155 }  ,draw opacity=1 ]   (154.87,164.6) -- (154.73,174.97) ;
\draw [color={rgb, 255:red, 155; green, 155; blue, 155 }  ,draw opacity=1 ][line width=0.75]    (180.2,164.6) -- (180.07,174.97) ;
\draw [color={rgb, 255:red, 155; green, 155; blue, 155 }  ,draw opacity=1 ]   (175.4,164.73) -- (175.27,175.1) ;
\draw [color={rgb, 255:red, 155; green, 155; blue, 155 }  ,draw opacity=1 ][line width=0.75]    (200.2,164.6) -- (200.07,174.97) ;
\draw [color={rgb, 255:red, 155; green, 155; blue, 155 }  ,draw opacity=1 ]   (195.4,164.73) -- (195.27,175.1) ;
\draw [color={rgb, 255:red, 155; green, 155; blue, 155 }  ,draw opacity=1 ][line width=0.75]    (190.07,164.47) -- (189.93,174.83) ;
\draw [color={rgb, 255:red, 155; green, 155; blue, 155 }  ,draw opacity=1 ]   (185.27,164.6) -- (185.13,174.97) ;
\draw [color={rgb, 255:red, 155; green, 155; blue, 155 }  ,draw opacity=1 ][line width=0.75]    (210.2,164.47) -- (210.07,174.83) ;
\draw [color={rgb, 255:red, 155; green, 155; blue, 155 }  ,draw opacity=1 ]   (205.4,164.6) -- (205.27,174.97) ;
\draw [color={rgb, 255:red, 155; green, 155; blue, 155 }  ,draw opacity=1 ]   (225.4,164.6) -- (225.27,174.97) ;
\draw [color={rgb, 255:red, 155; green, 155; blue, 155 }  ,draw opacity=1 ][line width=0.75]    (220.07,164.33) -- (219.93,174.7) ;
\draw [color={rgb, 255:red, 155; green, 155; blue, 155 }  ,draw opacity=1 ]   (215.27,164.47) -- (215.13,174.83) ;
\draw [color={rgb, 255:red, 155; green, 155; blue, 155 }  ,draw opacity=1 ][line width=0.75]    (240.4,164.6) -- (240.27,174.97) ;
\draw [color={rgb, 255:red, 155; green, 155; blue, 155 }  ,draw opacity=1 ]   (235.6,164.73) -- (235.47,175.1) ;
\draw [color={rgb, 255:red, 155; green, 155; blue, 155 }  ,draw opacity=1 ][line width=0.75]    (260.4,164.6) -- (260.27,174.97) ;
\draw [color={rgb, 255:red, 155; green, 155; blue, 155 }  ,draw opacity=1 ]   (255.6,164.73) -- (255.47,175.1) ;
\draw [color={rgb, 255:red, 155; green, 155; blue, 155 }  ,draw opacity=1 ][line width=0.75]    (250.27,164.47) -- (250.13,174.83) ;
\draw [color={rgb, 255:red, 155; green, 155; blue, 155 }  ,draw opacity=1 ]   (245.47,164.6) -- (245.33,174.97) ;
\draw [color={rgb, 255:red, 155; green, 155; blue, 155 }  ,draw opacity=1 ][line width=0.75]    (270.4,164.47) -- (270.27,174.83) ;
\draw [color={rgb, 255:red, 155; green, 155; blue, 155 }  ,draw opacity=1 ]   (265.6,164.6) -- (265.47,174.97) ;
\draw [color={rgb, 255:red, 155; green, 155; blue, 155 }  ,draw opacity=1 ]   (285.6,164.6) -- (285.47,174.97) ;
\draw [color={rgb, 255:red, 155; green, 155; blue, 155 }  ,draw opacity=1 ][line width=0.75]    (280.27,164.33) -- (280.13,174.7) ;
\draw [color={rgb, 255:red, 155; green, 155; blue, 155 }  ,draw opacity=1 ]   (275.47,164.47) -- (275.33,174.83) ;
\draw [color={rgb, 255:red, 155; green, 155; blue, 155 }  ,draw opacity=1 ][line width=0.75]    (300.53,164.6) -- (300.4,174.97) ;
\draw [color={rgb, 255:red, 155; green, 155; blue, 155 }  ,draw opacity=1 ]   (295.73,164.73) -- (295.6,175.1) ;
\draw [color={rgb, 255:red, 155; green, 155; blue, 155 }  ,draw opacity=1 ][line width=0.75]    (320.53,164.6) -- (320.4,174.97) ;
\draw [color={rgb, 255:red, 155; green, 155; blue, 155 }  ,draw opacity=1 ]   (315.73,164.73) -- (315.6,175.1) ;
\draw [color={rgb, 255:red, 155; green, 155; blue, 155 }  ,draw opacity=1 ][line width=0.75]    (310.4,164.47) -- (310.27,174.83) ;
\draw [color={rgb, 255:red, 155; green, 155; blue, 155 }  ,draw opacity=1 ]   (305.6,164.6) -- (305.47,174.97) ;
\draw [color={rgb, 255:red, 155; green, 155; blue, 155 }  ,draw opacity=1 ][line width=0.75]    (330.53,164.47) -- (330.4,174.83) ;
\draw [color={rgb, 255:red, 155; green, 155; blue, 155 }  ,draw opacity=1 ]   (325.73,164.6) -- (325.6,174.97) ;
\draw [color={rgb, 255:red, 155; green, 155; blue, 155 }  ,draw opacity=1 ]   (345.73,164.6) -- (345.6,174.97) ;
\draw [color={rgb, 255:red, 155; green, 155; blue, 155 }  ,draw opacity=1 ][line width=0.75]    (340.4,164.33) -- (340.27,174.7) ;
\draw [color={rgb, 255:red, 155; green, 155; blue, 155 }  ,draw opacity=1 ]   (335.6,164.47) -- (335.47,174.83) ;
\draw [color={rgb, 255:red, 155; green, 155; blue, 155 }  ,draw opacity=1 ][line width=0.75]    (360,164.33) -- (359.87,174.7) ;
\draw [color={rgb, 255:red, 155; green, 155; blue, 155 }  ,draw opacity=1 ]   (355.2,164.47) -- (355.07,174.83) ;
\draw [color={rgb, 255:red, 155; green, 155; blue, 155 }  ,draw opacity=1 ][line width=0.75]    (380,164.33) -- (379.87,174.7) ;
\draw [color={rgb, 255:red, 155; green, 155; blue, 155 }  ,draw opacity=1 ]   (375.2,164.47) -- (375.07,174.83) ;
\draw [color={rgb, 255:red, 155; green, 155; blue, 155 }  ,draw opacity=1 ][line width=0.75]    (369.87,164.2) -- (369.73,174.57) ;
\draw [color={rgb, 255:red, 155; green, 155; blue, 155 }  ,draw opacity=1 ]   (365.07,164.33) -- (364.93,174.7) ;
\draw [color={rgb, 255:red, 155; green, 155; blue, 155 }  ,draw opacity=1 ][line width=0.75]    (390,164.2) -- (389.87,174.57) ;
\draw [color={rgb, 255:red, 155; green, 155; blue, 155 }  ,draw opacity=1 ]   (385.2,164.33) -- (385.07,174.7) ;
\draw [color={rgb, 255:red, 155; green, 155; blue, 155 }  ,draw opacity=1 ]   (405.2,164.33) -- (405.07,174.7) ;
\draw [color={rgb, 255:red, 155; green, 155; blue, 155 }  ,draw opacity=1 ][line width=0.75]    (399.87,164.07) -- (399.73,174.43) ;
\draw [color={rgb, 255:red, 155; green, 155; blue, 155 }  ,draw opacity=1 ]   (395.07,164.2) -- (394.93,174.57) ;
\draw [color={rgb, 255:red, 155; green, 155; blue, 155 }  ,draw opacity=1 ][line width=0.75]    (420,164.33) -- (419.87,174.7) ;
\draw [color={rgb, 255:red, 155; green, 155; blue, 155 }  ,draw opacity=1 ]   (415.2,164.47) -- (415.07,174.83) ;
\draw [color={rgb, 255:red, 155; green, 155; blue, 155 }  ,draw opacity=1 ][line width=0.75]    (440,164.33) -- (439.87,174.7) ;
\draw [color={rgb, 255:red, 155; green, 155; blue, 155 }  ,draw opacity=1 ]   (435.2,164.47) -- (435.07,174.83) ;
\draw [color={rgb, 255:red, 155; green, 155; blue, 155 }  ,draw opacity=1 ][line width=0.75]    (429.87,164.2) -- (429.73,174.57) ;
\draw [color={rgb, 255:red, 155; green, 155; blue, 155 }  ,draw opacity=1 ]   (425.07,164.33) -- (424.93,174.7) ;
\draw [color={rgb, 255:red, 155; green, 155; blue, 155 }  ,draw opacity=1 ][line width=0.75]    (450,164.2) -- (449.87,174.57) ;
\draw [color={rgb, 255:red, 155; green, 155; blue, 155 }  ,draw opacity=1 ]   (445.2,164.33) -- (445.07,174.7) ;
\draw [color={rgb, 255:red, 155; green, 155; blue, 155 }  ,draw opacity=1 ]   (465.2,164.33) -- (465.07,174.7) ;
\draw [color={rgb, 255:red, 155; green, 155; blue, 155 }  ,draw opacity=1 ][line width=0.75]    (459.87,164.07) -- (459.73,174.43) ;
\draw [color={rgb, 255:red, 155; green, 155; blue, 155 }  ,draw opacity=1 ]   (455.07,164.2) -- (454.93,174.57) ;
\draw [color={rgb, 255:red, 155; green, 155; blue, 155 }  ,draw opacity=1 ][line width=0.75]    (480.13,164.33) -- (480,174.7) ;
\draw [color={rgb, 255:red, 155; green, 155; blue, 155 }  ,draw opacity=1 ]   (475.33,164.47) -- (475.2,174.83) ;
\draw [color={rgb, 255:red, 155; green, 155; blue, 155 }  ,draw opacity=1 ][line width=0.75]    (500.13,164.33) -- (500,174.7) ;
\draw [color={rgb, 255:red, 155; green, 155; blue, 155 }  ,draw opacity=1 ]   (495.33,164.47) -- (495.2,174.83) ;
\draw [color={rgb, 255:red, 155; green, 155; blue, 155 }  ,draw opacity=1 ][line width=0.75]    (490,164.2) -- (489.87,174.57) ;
\draw [color={rgb, 255:red, 155; green, 155; blue, 155 }  ,draw opacity=1 ]   (485.2,164.33) -- (485.07,174.7) ;
\draw [color={rgb, 255:red, 155; green, 155; blue, 155 }  ,draw opacity=1 ][line width=0.75]    (510.13,164.2) -- (510,174.57) ;
\draw [color={rgb, 255:red, 155; green, 155; blue, 155 }  ,draw opacity=1 ]   (505.33,164.33) -- (505.2,174.7) ;
\draw [color={rgb, 255:red, 155; green, 155; blue, 155 }  ,draw opacity=1 ]   (525.33,164.33) -- (525.2,174.7) ;
\draw [color={rgb, 255:red, 155; green, 155; blue, 155 }  ,draw opacity=1 ][line width=0.75]    (520,164.07) -- (519.87,174.43) ;
\draw [color={rgb, 255:red, 155; green, 155; blue, 155 }  ,draw opacity=1 ]   (515.2,164.2) -- (515.07,174.57) ;

\draw (733,51.4) node [anchor=north west][inner sep=0.75pt]    {$$};
\draw (41,185.7) node [anchor=north west][inner sep=0.75pt]  [font=\tiny] [align=left] {{\footnotesize year 0}};
\draw (101,185.7) node [anchor=north west][inner sep=0.75pt]  [font=\tiny] [align=left] {{\footnotesize year 1}};
\draw (161,185.7) node [anchor=north west][inner sep=0.75pt]  [font=\tiny] [align=left] {{\footnotesize year 2}};
\draw (223,184.7) node [anchor=north west][inner sep=0.75pt]  [font=\tiny] [align=left] {{\footnotesize year 3}};
\draw (283,184.7) node [anchor=north west][inner sep=0.75pt]  [font=\tiny] [align=left] {{\footnotesize year 4}};
\draw (343,184.7) node [anchor=north west][inner sep=0.75pt]  [font=\tiny] [align=left] {{\footnotesize year 5}};
\draw (404,184.7) node [anchor=north west][inner sep=0.75pt]  [font=\tiny] [align=left] {{\footnotesize year 6}};
\draw (464,184.7) node [anchor=north west][inner sep=0.75pt]  [font=\tiny] [align=left] {{\footnotesize year 7}};
\draw (524,184.7) node [anchor=north west][inner sep=0.75pt]  [font=\tiny] [align=left] {{\footnotesize year 8}};

\end{tikzpicture}
\label{fig:rollingSample}
\end{figure}

The optimization model to decide the asset allocation weights is run once a year (in January), based on 5-year lookback periods.  In essence, the optimization is based on a sequence of overlapping windows as shown in Figure (\ref{fig:rollingSample}).  Hence,  the first optimization is based on data from the the January 2003-December 2007 period. And the merits of this asset-class selection (out-of-sample performance) are evaluated a year later, in January 2009 (backtesting).  Then, a second optimization is run based on the January 2004-December 2008 period data, and its performance is evaluated, this time,  in January 2010.  This backtesting process is repeated until reaching the January 2017-December 2021 period.  Note that this last weight selection is tested over a shorter time-window (January 2022-June 2022).  Also, each optimization problem is solved for several CVaR limits, ranging from  7.5\%   to  30\%,   to capture the preferences of investors with different risk-tolerance  levels.
Additionally, given that the proposed procedure is non-deterministic (mainly because of the synthetic nature of the returns generated when using CTGAN) each optimization is run 5 times for each  CVaR tolerance  level  
 ($\Lambda $) .  This allows us to test the stability of the results.  Finally, note that in the cases with no features
 the density vector $\bm{\pi}\in\mathbb{R}^m$ is defined as $\pi_j = \frac{1}{m}$.
  for $j \in \{1,...,m\}$.

In summary, the testing strategy is really a sequence of fourteen backtesting exercises starting in January 2009, and performed annually, until January 2022, plus, one final test done in July 2022 (based on a 6-month window, January 2022-June 2022).  This process is summarized in a schematic fashion in Figure \ref{fig:backtesting}.

\subsection{   Performance  metrics   }

Comparing the performance of investment strategies over long time-horizons (an asset  allocation scheme is ultimately an investment strategy)  is a multidimensional exercise  that should take into account  several factors, namely, returns, risk, level of trading, degree of portfolio diversification, etc.  

To this end, we consider four metrics (figures of merits) to carry out our comparisons.  These comparisons are based on the performance (determined via backtesting) over the January 2008-June 2022 period, in all, 14.5 years.

We  consider the following metrics:

\begin{enumerate}
    \item\textbf{Returns: } Returns constitute the quintessential performance  yardstick.  Since we are dealing with a medium- to long-term horizon investor, the cumulative return over this 14.5-year period, expressed in annualized form, is the best metric to assess returns.

    \item\textbf{Risk: } Since we have formulated the optimization problem based on a CVaR constraint, it makes sense to check the  CVaR ex  post.  A gross violation of the CVaR limit should raise concerns regarding the benefits of the strategy.

    \item\textbf{Transaction costs: }   Notwithstanding the fact that rebalancing is done once a year, transaction costs, at least in theory, could be significant.  Portfolio rotation is a good proxy to assess the impact of transaction costs (which, if excessive, could negatively affect returns).  The level of portfolio rotation, on an annual basis, can be expressed as

    \begin{align}
    \label{eqn:rotation}
    \textsc{rotation} = \frac{\sum_{t=2}^{14}  \sum_{i=1}^{10} |w_{i,t} - w_{i,t-1}|}{14}
    \end{align}
where the  $ \omega $'s are the asset allocation weights.  A static portfolio results in a value equal 0; increasing values of this metric are associated with increasing levels  of portfolio rotation.

    \item\textbf{Diversification: }  Most investors aim at having a diversified portfolio.  (Recall that a frequent criticism to the conventional MV-approach is that it often yields corner solutions  based on portfolios heavily concentrated on a few assets.)   To measure the degree of diversificatio, we follow \citet{pagnocelli2022}, and   rely on the complementary Herfindahl–Hirschman (HH)  Index.  A value of  0 for the index reflects a portfolio concentrated on a single asset. On the other hand, a value  approaching 1 corresponds to a fully diversified portfolio (all assets  share the same weight).
\end{enumerate}

\begin{figure}[H]
\centering
\tbl{Overview of Backtesting Method}

\tikzset{every picture/.style={line width=0.75pt}} 

\begin{tikzpicture}[x=0.75pt,y=0.75pt,yscale=-1,xscale=1]

\draw   (244,149) -- (418,149) -- (418,196.5) -- (244,196.5) -- cycle ;
\draw   (176,406.5) -- (500,406.5) -- (500,554.5) -- (175,554.5) -- cycle ;
\draw    (332,200) -- (332,225.5) ;
\draw [shift={(332,227.5)}, rotate = 270] [color={rgb, 255:red, 0; green, 0; blue, 0 }  ][line width=0.75]    (10.93,-3.29) .. controls (6.95,-1.4) and (3.31,-0.3) .. (0,0) .. controls (3.31,0.3) and (6.95,1.4) .. (10.93,3.29)   ;
\draw    (333,360) -- (333,399) ;
\draw [shift={(333,401)}, rotate = 270] [color={rgb, 255:red, 0; green, 0; blue, 0 }  ][line width=0.75]    (10.93,-3.29) .. controls (6.95,-1.4) and (3.31,-0.3) .. (0,0) .. controls (3.31,0.3) and (6.95,1.4) .. (10.93,3.29)   ;
\draw   (275,31) -- (381,31) -- (381,97) -- (275,97) -- cycle ;
\draw    (331,98) -- (331,141) ;
\draw [shift={(331,143)}, rotate = 270] [color={rgb, 255:red, 0; green, 0; blue, 0 }  ][line width=0.75]    (10.93,-3.29) .. controls (6.95,-1.4) and (3.31,-0.3) .. (0,0) .. controls (3.31,0.3) and (6.95,1.4) .. (10.93,3.29)   ;
\draw  [color={rgb, 255:red, 255; green, 255; blue, 255 }  ,draw opacity=1 ] (54,218) -- (228,218) -- (228,265.5) -- (54,265.5) -- cycle ;
\draw   (188,234.5) -- (475,234.5) -- (475,353.5) -- (188,353.5) -- cycle ;

\draw (296.5,166.9) node [anchor=north west][inner sep=0.75pt]    {$\boldsymbol{f}_{t} \ \in \mathbb{R}{^{l}} \ $};
\draw (190,243) node [anchor=north west][inner sep=0.1]  [font=\scriptsize] [align=left] {Compute the Euclidean distance between the normalized \\ present day features and the sample features};
\draw (733,51.4) node [anchor=north west][inner sep=0.75pt]    {$$};
\draw (201,276.4) node [anchor=north west][inner sep=0.75pt]    {$d(\boldsymbol{f}_{1} ,\boldsymbol{f}_{2}) \ =\ \sqrt{(\boldsymbol{f}_{1} \ -\ \boldsymbol{f}_{2})^{T}(\boldsymbol{f}_{1} \ -\ \boldsymbol{f}_{2})}$};
\draw (230,440) node [anchor=north west][inner sep=0.75pt]    {$x_{Gw/oF} \coloneq  \textsc{Optimization}\left(\mathbf{R}_{t}^{s} ,\frac{\boldsymbol{1}}{m}\right)$};
\draw (185,414) node [anchor=north west][inner sep=0.75pt]  [font=\scriptsize] [align=left] {Run the optimization for different risk-tolerance levels with \\ the appropriate weights for each scenario};
\draw (230,13) node [anchor=north west][inner sep=0.75pt]  [font=\scriptsize] [align=left] {Start with a set of returns and features};
\draw (289,67.4) node [anchor=north west][inner sep=0.75pt]    {$\mathbf{F}^{v} \in \ \mathbb{R}^{l,\ m_{v}} \ $};
\draw (333,107) node [anchor=north west][inner sep=0.75pt]  [font=\scriptsize] [align=left] {Loop over all rebalance \\ \ \ \ \ \ \ \ days t$\displaystyle \in T$};
\draw (252,155) node [anchor=north west][inner sep=0.75pt]  [font=\scriptsize] [align=left] {Obtain the present day features};
\draw (244,312.4) node [anchor=north west][inner sep=0.75pt]    {$\boldsymbol{\pi _{f_t}} \ =\ \frac{\boldsymbol{d_{f_t}^{-1}}}{\boldsymbol{1^{T} d_{f_t}^{-1}}}$};
\draw (230,468) node [anchor=north west][inner sep=0.75pt]    {$x_{Hw/oF} \coloneq  \textsc{Optimization}\left(\mathbf{R}_{t}^{h} ,\frac{\boldsymbol{1}}{m}\right)$};
\draw (288,40.4) node [anchor=north west][inner sep=0.75pt]    {$\mathbf{R}^{v} \in \ \mathbb{R}^{n,\ m_{v}} \ $};
\draw (240,496) node [anchor=north west][inner sep=0.75pt]     {$x_{HwF} \coloneq  \textsc{Optimization}\left(\mathbf{R}_{t}^{h} ,\boldsymbol{\pi _{f_t}}\right)$};
\draw (240,530) node [anchor=north west][inner sep=0.75pt]    {$x_{GwF} \coloneq \textsc{Optimization}\left(\mathbf{R}_{t}^{s} ,\boldsymbol{\pi _{f_t}}\right)$};

\end{tikzpicture}

\label{fig:backtesting}

\end{figure}

\subsection{Performance  comparison }
For comparison purposes, all numerical experiments were run on a MacBook Pro 14 with an M1 Pro chip and 16 GB of RAM. All the strategies were run without the use of a dedicated GPU to be able to perform a fair comparison across strategies.

The strategies were backtested using a 5-year window of daily historical scenarios as input. In the case of  the CTGAN-based strategies(Gw/oF and GwF) all the 5-year window historical scenarios were used as input for the Data Generating Process, then, a sample of 500 synthetic scenarios were used to solve the optimization problem.  In the case of the historical-based strategies (Hw/oF and HwF) the inputs were a sub-sample of 500 historical scenarios which were used to solve the optimization problem. In the case of the EW strategy there is no such input or sub-sampling since the strategy does not dependent on any scenarios: the weights are always the same and identical.

Regarding the historical-based strategies (Hw/oF and HwF) the running time was on average $0.001$ seconds per rebalance cycle. The running time for the CTGAN based strategies (Gw/oF and GwF) was on average $203.5$ seconds per rebalance cycle. Given that all strategies were run using only CPU and not GPU-accelerated hardware the CTGAN based strategies were slower to run given the greater number of operations used to train a GAN-based architecture.

Figure \ref{fig:figures} shows the values of all relevant metrics.

We start with the  returns.  First, the benefits of including features (contextual information) in the optimization process are evident: both, the GwF and HwF approaches, outperform by far their non-features counterparties. The difference in performance is more manifest as the CVaR limit increases. Intuitively, this makes sense:  stricter risk limits tend to push the solutions towards cash-based instruments, which, in turn, exhibit returns that are less dependent on the economic environment, and thus, the benefits of the information-content embedded in the features is diminished.  Note also that all strategies (except for the EW) deliver, more or less,  monotonically increasing returns as the CVaR limit is relaxed.  
Additionally, it is worth mentioning that a  naive visual inspection might suggest that GwF only outperforms HwF by a fairly small margin.  Take the case of CVaR = 0.25, for example; the difference between 16.78\% and 15.65\% might appear as innocuous.  Over a 14.5-year period, however, it is significant. More clearly: an investor who contributed \$ 100 to the GwF  strategy initially,  will end up with \$ 948; the investor who adopted the HwF strategy, will end up with only \$ 823.  
We should  be careful not to jump to conclusions regarding the merits of including features in asset allocation problems. However, our results strongly suggest that the benefits of incorporating features to the optimization framework can  be substantial. Finally, the EW  strategy clearly underperforms compared to all other strategies.

We now turn  to the CVaR (ex post).  Again, the benefits of including features are clear as they always decrease the risk compared to the non-features options.  Also noticeably, including features (see HwF and GwF) always yields solutions that never violate the CVaR limit established ex ante.  It might seem surprising that the CVaR-ex post value does not increase monotonically as the CVaR limit (actually $\Lambda$ based on the notation used in (\ref{eqn:assetalloc_optproblem}), increases, especially in the GwF and HwF cases.  We attribute this situation to the fact that the  CVaR-restriction  was probably not active when the optimization reached a solution.

In terms of diversification (HH Index), all in all, all strategies display fairly similar diversification levels.  Two comments are in order.  First, relaxing the risk limit (higher CVaR) naturally results in lower diversification as the portfolios tend to move to higher-yielding assets, which are, in general, riskier.  And second, it might appear that the overall diversification level is low (values of the HH Index below 0.20 in most cases).  That sentiment,  however, would be misplaced: these are portfolios made up, not of individual assets, but indices, and thus, they are inherently  highly diversified. 

Lastly, we examine trading expenses.  It might be difficult from Table \ref{table:rotation}, Rotation, to gauge its impact on returns. To actually estimate rigorously the potential impact of trading expenses on returns, in all cases, we proceed as follows.   Table \ref{table:expenses} shows for different asset classes (based on some commonly traded and liquid ETFs), representative bid-ask spreads. This information,  in combination with the rotation levels shown in Table \ref{table:rotation}, can be used to estimate the trading expenses on a per annum basis (shown in Table \ref{table:trxcost}).  Finally, Table \ref{table:netreturns} shows the returns after correcting  for trading expenses.  A comparison between these returns and those shown  in Table \ref{table:returns} proves that trading expenses have no significant impact on returns.

In summary, all things considered, features-based strategies outperform their versions with no features, and, more important,  GwF clearly outperforms HwF, most evidently in terms of returns, the variable investors care the most.  The EW strategy, which had done surprisingly well against MV-based portfolios, emerges as the clear loser, by far.

\begin{figure}[H]
\centering
\caption{Key Metrics for All Strategies}
\hspace{-40pt} 
\begin{minipage}{0.4\textwidth}
    \centering
    \small
    \textbf{ (a) Annualized Returns}
    \begin{tabular}{cccccc}\hline
        CVaR   &   Gw/oF   &    GwF   &   Hw/oF    &   HwF     &   EW  \\ \hline
        0.075  &  12.54\%  & 13.50\%  &   12.90\%  &  12.74\%  &  7.89\%  \\
        0.1    &  11.96\%  & 13.30\%  &   12.73\%  &  12.98\%  &  7.89\%  \\
        0.125  &  12.46\%  & 14.94\%  &   13.04\%  &  13.67\%  &  7.89\%  \\
        0.15   &  13.84\%  & 15.43\%  &   13.20\%  &  14.03\%  &  7.89\%  \\
        0.175  &  12.95\%  & 15.18\%  &   14.04\%  &  14.08\%  &  7.89\%  \\
        0.2    &  13.02\%  & 15.21\%  &   13.57\%  &  14.71\%  &  7.89\%  \\
        0.225  &  12.51\%  & 16.22\%  &   13.26\%  &  15.20\%  &  7.89\%  \\
        0.25   &  13.19\%  & 16.78\%  &   13.31\%  &  15.65\%  &  7.89\%  \\
        0.275  &  13.60\%  & 17.36\%  &   13.59\%  &  16.45\%  &  7.89\%  \\
        0.3    &  13.87\%  & 17.77\%  &   14.90\%  &  16.64\%  &  7.89\%  \\
    \end{tabular}
    \label{table:returns}
\end{minipage}
\hfill 
\begin{minipage}{0.45\textwidth}
    \centering
    \small
    \textbf{ (b) CVaR Ex-post}
    \begin{tabular}{cccccc}\hline
        CVaR   &   Gw/oF      &    GwF   &   Hw/oF   &   HwF    &   EW  \\ \hline
        0.075  &  10.20\%      &   4.56\%    &   7.34\%    & 6.58\%     &   5.33\% \\
        0.10  &  10.75\%      &   7.31\%    &   8.76\%    & 5.51\%     &   5.33\% \\
        0.125  &  10.01\%      &   4.68\%    &   8.65\%    & 4.21\%     &   5.33\% \\
        0.15  &  8.62\%       &   4.71\%    &   8.59\%    & 3.87\%     &   5.33\% \\
        0.175  &  10.21\%      &   5.90\%    &   7.22\%    & 3.42\%     &   5.33\% \\
        0.20  &  10.29\%      &   5.41\%    &   8.60\%    & 3.49\%     &   5.33\% \\
        0.225  &  10.77\%      &   5.11\%    &   9.97\%    & 3.62\%     &   5.33\% \\
        0.25  &  10.15\%      &   4.12\%    &   9.98\%    & 3.89\%     &   5.33\% \\
        0.275  &  10.71\%     &   6.18\%   &   9.98\%    & 4.24\%     &   5.33\% \\
        0.3    &  10.41\%     &   4.67\%   &   7.61\%    & 4.71\%     &   5.33\% 
    \end{tabular}
    \label{table:cvar}
\end{minipage}

\bigskip

\hspace{-20pt} 
\begin{minipage}{0.4\textwidth}
    \centering
    \small
    \textbf{ (c) HH Index}
    \begin{tabular}{cccccc}\hline
        CVaR  &   Gw/oF   & GwF   &   Hw/oF  &   HwF   &   EW  \\ \hline
        0.075 &    0.18  &   0.25 &     0.17 &   0.19  & 1  \\
        0.10 &    0.15  &   0.23 &     0.18 &   0.18  & 1  \\
        0.125 &    0.17  &   0.20 &     0.19 &   0.18  & 1  \\
        0.15 &    0.18  &   0.18 &     0.19 &   0.16  & 1  \\
        0.175 &    0.17  &   0.16 &     0.20 &   0.16  & 1  \\
        0.20 &    0.18  &   0.15 &     0.19 &   0.17  & 1  \\
        0.225 &    0.17  &   0.18 &     0.18 &   0.18  & 1  \\
        0.25 &    0.17  &   0.12 &     0.18 &   0.18  & 1 \\
        0.275 &    0.16  &   0.10 &     0.17 &   0.18  & 1 \\
        0.3   &    0.16  &   0.12 &     0.17 &   0.17  & 1 
    \end{tabular}
    \label{table:hhindex}
\end{minipage}
\hfill
\begin{minipage}{0.45\textwidth}
    \centering
    \small
    \textbf{ (d) Rotation}
    \begin{tabular}{cccccc}\hline
        CVaR  & Gw/oF  & GwF   &   Hw/oF  &   HwF   &   EW  \\ \hline
        0.075 & 14.35   & 28.91   & 9.38   & 24.53   & 0  \\
        0.10  & 13.98   & 34.76   & 10.15  & 26.96   & 0  \\
        0.125 & 13.92   & 26.23   & 10.43  & 25.95   & 0  \\
        0.15  & 14.20   & 26.60   & 10.49  & 28.48   & 0  \\
        0.175 & 14.95   & 26.34   & 10.25  & 26.80   & 0  \\
        0.20  & 14.07   & 31.25   & 10.95  & 24.43   & 0  \\
        0.225 & 15.46   & 24.12   & 12.01  & 23.76   & 0  \\
        0.25  & 15.19   & 39.00   & 12.15  & 21.53   & 0 \\
        0.275 & 14.42   & 19.78   & 12.61  & 19.00   & 0 \\
        0.3   & 15.03   & 19.45   & 10.91  & 17.01   & 0 
    \end{tabular}
    \label{table:rotation}
\end{minipage}

\label{fig:figures}
\end{figure}

\begin{figure}[H]
    \begin{table}[H]
    \tbl{Trading Expenses by Asset Class}
        {\begin{tabular}{cccccc}\hline
            Asset Class   &   Selected ETF   &    Average 30 Day Bid-Ask Spread (Basis Points) \\ \hline
            US equities              & SPY US   & 0.36 \\
            US equities tech         & QQQ US   & 0.52  \\
            Global equities          & VT US    & 0.54 \\
            EM equities              & EEM US   & 2.69 \\
            US high yield                    & HYG US   & 1.35 \\
            US inv. grade                    & LQD US   & 0.96 \\
            EM debt                  & PCY US   & 5.66 \\
            Commodities              & COMT US  & 14.1 \\
            Long term treasuries     & TLT US   & 1.03 \\
            Short term treasuries    & BIL US   & 1.25 \\

        \end{tabular}}
    \label{table:expenses}
    \end{table}
\end{figure}

\begin{figure}[H]
    \begin{table}[H]
    \tbl{Annualized Transaction Expenses \\(Basis Points) }
        {\begin{tabular}{cccccc}\hline
            CVaR   &   Gw/oF       &    GwF        &   Hw/oF   &   HwF    &   EW  \\ \hline
            0.075  &  0.54      &   1.32    &   0.19    & 1.52     &   0 \\
            0.10  &  0.43      &   1.50    &   0.23    & 1.72     &   0 \\
            0.125 &  0.44      &   1.20    &   0.23    & 1.64     &   0 \\
            0.15  &  0.47      &   1.61    &   0.23    & 1.82     &   0 \\
            0.175 &  0.53      &   1.31    &   0.24    & 1.76     &   0 \\
            0.20  &  0.47      &   1.46    &   0.25    & 1.56     &   0 \\
            0.225 &  0.49      &   0.99    &   0.28    & 1.49     &   0 \\
            0.25  &  0.53      &   1.51    &   0.28    & 1.30     &   0 \\
            0.275 &  0.45      &   0.80    &   0.30    & 1.07     &   0 \\
            0.30  &  0.45      &   0.85    &   0.27    & 0.93     &   0 \\
        \end{tabular}}
    \label{table:trxcost}
    \end{table}
\end{figure}
\hfill
\begin{figure}[H]
    \begin{table}[H]
    \tbl{Annualized Returns \\ (Net of Transaction Expenses)}
        {\begin{tabular}{cccccc}\hline
            CVaR   &   Gw/oF  & GwF   &   Hw/oF  &   HwF   &   EW  \\ \hline
            0.075  &  12.53\%  & 13.49\%  &   12.90\%  &  12.72\%  &  7.89\%  \\
            0.1    &  11.96\%  & 13.28\%  &   12.73\%  &  12.96\%  &  7.89\%  \\
            0.125  &  12.46\%  & 14.93\%  &   13.04\%  &  13.65\%  &  7.89\%  \\
            0.15   &  13.84\%  & 15.41\%  &   13.20\%  &  14.01\%  &  7.89\%  \\
            0.175  &  12.94\%  & 15.17\%  &   14.04\%  &  14.06\%  &  7.89\%  \\
            0.2    &  13.02\%  & 15.20\%  &   13.57\%  &  14.69\%  &  7.89\%  \\
            0.225  &  12.51\%  & 16.21\%  &   13.26\%  &  15.19\%  &  7.89\%  \\
            0.25   &  13.18\%  & 16.76\%  &   13.31\%  &  15.64\%  &  7.89\%  \\
            0.275  &  13.60\%  & 17.35\%  &   13.59\%  &  16.44\%  &  7.89\%  \\
            0.3    &  13.87\%  & 17.76\%  &   14.90\%  &  16.63\%  &  7.89\%  \\
        \end{tabular}}
    \label{table:netreturns}
    \end{table}
\end{figure}


\subsection{Discussion of results and some considerations regarding potential statistical biases}

Broadly speaking, presenting a model that outperforms a benchmark is not an insurmountable task. In this case, we have presented a model (strategy or method) that both generates realistic  synthetic data and delivers satisfactory out-of-sample performance. Given this situation, reasonable readers might ask themselves: How well would the model proposed perform under circumstances different from those described in the example selected by the authors? Did the authors fine-tune the value of some critical parameters in order to present their results in the best possible light? Do the results suffer from any form of selection bias? Overfitting and other statistical biases are common problems that affect many novel strategies and methods. Is there any indication of overfitting in this case? The following considerations are aimed at mitigating these concerns.

First, and in reference to a potential model selection bias.  The synthetic data generation approach we have presented is based on a Modified CTGAN model. We also considered two other potential choices for synthetic data generation, and we discarded them both. One was the NORTA (Normal to Anything) algorithm, a method based on the Gaussian copula that can generate vectors given a certain interdependence structure. This method has been successfully used in some financial applications  \citep{pagnocelli2022}  and delivered good out-of-sample performance.  Unfortunately, this algorithm requires to perform the Cholesky decomposition of the correlation matrix, a computational exercise of order $O(n^3)$, which makes the process computationally very expensive when one has many indices (ten in our case) combined with several features (eight in our example). In short, computationally speaking, NORTA was no match for CTGAN. A second alternative we considered, and decided not to explore, was the CopulaGAN method, a variation of GAN in which a copula-based algorithm is used to preprocess the data before applying the GAN model. This method is relatively new, and there is a lack of both academic literature and practical experience to make a strong case for CopulaGAN versus CTGAN. Hence, we also decided not to test it in our study.

Second, in reference to overfitting and selection bias. Like most neural networks, CTGAN relies on a set of hyperparameters. To avoid overfitting, we excluded any hyperparameter-tuning process. In fact, we maintained the number of layers, dimensions, and architecture of the CTGAN model proposed in \citep{leixu2019}, which also matched the default values of the model library. The only parameters that were modified were the learning rate, reduced to $10^{-4}$ from $2\times10^{-4}$, and the number of epochs (increased from 300 to 1500). These values proved to yield stable results across all  runs. It is important to mention that smoothing the learning rate and increasing the number of epochs does not affect the optimal solution, but guarantees a closer convergence at the expense of a higher (but still tolerable) computational cost. In the case of the remaining components of our proposed Synthetic Data Generation Process, namely TSNE, PCA, and HDBSCAN, we also decided not to tune any parameters, relying instead on the original implementations as they are.

Third, in reference to the lookback period (five years) and rebalancing period (one year), we did not test other lookback periods. However, previous experience suggests that the optimal length for lookback periods should be between three and five years  (\citet{GUTIERREZ2019134}). A period less than three years does not offer enough variability to capture key elements of the DGP, while periods longer than five years bring the risk of sampling from a "different universe" as financial markets are subject to exogenous conditions (e.g., regulation) that change over time. In other words, sampling returns from too far in the past could bring elements into the modeling process that may not reflect current market dynamics. Additionally, we did not test rebalancing periods different than one year. Rebalancing periods much shorter than one year probably do not make sense in the context of passive investment, which is the philosophy behind the investment approach we are advocating. And from a practical point of view, most investors would not entertain a rebalancing period less frequent than once a year since in general people evaluate their investment priorities on a yearly basis.

In brief, we hope that these additional explanations will be helpful in evaluating the relevance of our results and dispelling any major concerns related to potential biases.

\section{Conclusions}

Several conclusions emerge from this study. The most important is that the synthetic data generating approach suggested (based on a Modified CTGAN method enhanced with contextual information) seems very promising. First, it generates data (in this case returns) that capture well the essential character of historical data. And second, such data, when used in conjunction with the CVaR-based optimization framework described in (\ref{eqn:general_discrete_problem}), yields portfolios with satisfactory out-of-sample performance.

Additionally, the example also emphasizes the benefits of incorporating contextual information.  Recall that both, the GwF and HwF methods outperformed clearly their non-features counterparties.  Also, the fact that the GwF approach outperformed the HwF approach, highlights both, the shortcomings of methods based only on historical data, and the relevance of including scenarios that even though have not occurred, are “feasible”, given the nature of the historical data.  This element, we think, is critical to achieve a good out-of-sample performance.

However, notwithstanding the fact that the example presented captured a challenging period for the financial markets (subprime and COVID crises), and considered a broad set of assets (stocks, bonds, and commodities), the results should be interpreted with restraint.  That is, as an invitation to explore in more detail certain topics, rather than falling into the temptation of making absolute statements about the merits of the methods we have presented.  In fact, two topics that deserve further exploration are: (i) the benefits of using alternatives other than the different tenors of the yield curve as features, or, perhaps, using the yield curve in combination with other data (e.g., market volatility, liquidity indices, currency movements); and (ii) the use of the synthetic data generating method we proposed applied to financial variables other than returns, for example, bond default rates, or, exchange rates.  We leave these challenges for future research efforts.

\section*{Disclosure statement}

The authors report there are no competing interests to declare.

\section*{Data availability statement}

The code and data that support the findings of this study are openly available in GitHub at \url{https://github.com/chuma9615/ctgan-portfolio-research}, Historical data was obtained from Bloomberg. 

\bibliography{interactapasample}

\end{document}